\documentclass[aps,prx,showpacs,notitlepage,floatfix,onecolumn,superscriptaddress, nofootinbib, twocolumn]{revtex4-2}
\usepackage{graphicx} % Required for inserting images
\usepackage{amsmath,amssymb}
\usepackage{tikz}
\usetikzlibrary{shapes.misc}
\usetikzlibrary{positioning,fit}
\usetikzlibrary{calc}
\usepackage{braket}
\usepackage{amsthm}
\usepackage{float}
\makeatletter
\let\petz@original@newfloat\newfloat
\let\newfloat\newfloat@ltx
\makeatother
\usepackage{algorithm}
\makeatletter
\let\newfloat\petz@original@newfloat
\makeatother
\usepackage{algpseudocode}
\usepackage[colorlinks,linkcolor=blue,anchorcolor=blue,citecolor=blue,urlcolor=blue]{hyperref}
\def\tr{\mathrm{tr}}
\def\dim{\mathrm{dim}}
\def\E{\mathcal{E}}
\def\R{\mathcal{R}}
\def\H{\mathcal{H}}
\graphicspath{{figure/}}

\usepackage{tikz}
\usetikzlibrary{shapes.misc}
\usetikzlibrary{positioning,fit}
\usetikzlibrary{calc}

\tikzset{
  unitary/A label/.initial={},
  unitary/B label/.initial={},
  unitary/C label/.initial={},
  unitary/D label/.initial={},
  pics/unitary/.style={
    code={
      \node[rectangle, draw=black, fill=white!5, very thick,
        minimum width=1.6cm, minimum height=0.6cm] (-body) at (0, 0) {$U$};
      \draw[thick] ([xshift=-.5cm,yshift=0cm]-body.north)--++(0,0.3cm) coordinate (-C) node[above]  {\pgfkeysvalueof{/tikz/unitary/C label}};
      \draw[thick] ([xshift=.5cm,yshift=0cm]-body.north)--++(0,0.3cm) coordinate (-D) node[above] {\pgfkeysvalueof{/tikz/unitary/D label}};
      \draw[thick] ([xshift=-.5cm,yshift=0cm]-body.south)--++(0,-0.3cm) coordinate (-A) node[below]  {\pgfkeysvalueof{/tikz/unitary/A label}};
      \draw[thick] ([xshift=.5cm,yshift=0cm]-body.south)--++(0,-0.3cm) coordinate (-B) node[below]  {\pgfkeysvalueof{/tikz/unitary/B label}};
    }
  },
  pics/unitaryd/.style={
    code={
      \node[rectangle, draw=black, fill=white!5, very thick,
        minimum width=1.6cm, minimum height=0.6cm] (-body) at (0, 0) {$U^\dagger$};
    \draw[thick] ([xshift=-.5cm,yshift=0cm]-body.north)--++(0,0.3cm) coordinate (-A) node[above]  {\pgfkeysvalueof{/tikz/unitary/A label}};
      \draw[thick] ([xshift=.5cm,yshift=0cm]-body.north)--++(0,0.3cm) coordinate (-B) node[above] {\pgfkeysvalueof{/tikz/unitary/B label}};
      \draw[thick] ([xshift=-.5cm,yshift=0cm]-body.south)--++(0,-0.3cm) coordinate (-C) node[below]  {\pgfkeysvalueof{/tikz/unitary/C label}};
      \draw[thick] ([xshift=.5cm,yshift=0cm]-body.south)--++(0,-0.3cm) coordinate (-D) node[below]  {\pgfkeysvalueof{/tikz/unitary/D label}};
    }
  },
    pics/unitaryt/.style={
    code={
      \node[rectangle, draw=black, fill=white!5, very thick,
        minimum width=1.6cm, minimum height=0.6cm] (-body) at (0, 0) {$U^T$};
    \draw[thick] ([xshift=-.5cm,yshift=0cm]-body.north)--++(0,0.3cm) coordinate (-B) node[above]  {\pgfkeysvalueof{/tikz/unitary/B label}};
      \draw[thick] ([xshift=.5cm,yshift=0cm]-body.north)--++(0,0.3cm) coordinate (-A) node[above] {\pgfkeysvalueof{/tikz/unitary/A label}};
      \draw[thick] ([xshift=-.5cm,yshift=0cm]-body.south)--++(0,-0.3cm) coordinate (-D) node[below]  {\pgfkeysvalueof{/tikz/unitary/D label}};
      \draw[thick] ([xshift=.5cm,yshift=0cm]-body.south)--++(0,-0.3cm) coordinate (-C) node[below]  {\pgfkeysvalueof{/tikz/unitary/C label}};
    }
  },
  pics/unitarystar/.style={
    code={
      \node[rectangle, draw=black, fill=white!5, very thick,
        minimum width=1.6cm, minimum height=0.6cm] (-body) at (0, 0) {$U^*$};
      \draw[thick] ([xshift=-.5cm,yshift=0cm]-body.north)--++(0,0.3cm) coordinate (-D) node[above]  {\pgfkeysvalueof{/tikz/unitary/D label}};
      \draw[thick] ([xshift=.5cm,yshift=0cm]-body.north)--++(0,0.3cm) coordinate (-C) node[above] {\pgfkeysvalueof{/tikz/unitary/C label}};
      \draw[thick] ([xshift=-.5cm,yshift=0cm]-body.south)--++(0,-0.3cm) coordinate (-B) node[below]  {\pgfkeysvalueof{/tikz/unitary/B label}};
      \draw[thick] ([xshift=.5cm,yshift=0cm]-body.south)--++(0,-0.3cm) coordinate (-A) node[below]  {\pgfkeysvalueof{/tikz/unitary/A label}};
    }
  },
}

\tikzset{
  kraus/A label/.initial={},
  kraus/C label/.initial={},
  kraus/D label/.initial={},
  kraus/label/.initial={$\colE{I}$},
  pics/kraus/.style={
    code={
      \node[rectangle, draw=black, fill=white!5, very thick,
        minimum width=1.6cm, minimum height=0.6cm] (-body) at (0, 0) {\pgfkeysvalueof{/tikz/kraus/label}};
      \draw[thick] ([xshift=-.5cm,yshift=0cm]-body.north)--++(0,0.3cm) coordinate (-C) node[above]  {\pgfkeysvalueof{/tikz/kraus/C label}};
      \draw[thick] ([xshift=.5cm,yshift=0cm]-body.north)--++(0,0.3cm) coordinate (-D) node[above] {\pgfkeysvalueof{/tikz/kraus/D label}};
      \draw[thick] ([xshift=-.5cm,yshift=0cm]-body.south)--++(0,-0.3cm) coordinate (-A) node[below]  {\pgfkeysvalueof{/tikz/kraus/A label}};
    }
  },
  pics/krausstar/.style={
    code={
       \node[rectangle, draw=black, fill=white!5, very thick,
        minimum width=1.6cm, minimum height=0.6cm] (-body) at (0, 0) {\pgfkeysvalueof{/tikz/kraus/label}};
      \draw[thick] ([xshift=-.5cm,yshift=0cm]-body.north)--++(0,0.3cm) coordinate (-D) node[above]  {\pgfkeysvalueof{/tikz/kraus/D label}};
      \draw[thick] ([xshift=.5cm,yshift=0cm]-body.north)--++(0,0.3cm) coordinate (-C) node[above] {\pgfkeysvalueof{/tikz/kraus/C label}};
      \draw[thick] ([xshift=.5cm,yshift=0cm]-body.south)--++(0,-0.3cm) coordinate (-A) node[below]  {\pgfkeysvalueof{/tikz/kraus/A label}};
    }
  },
    pics/krausd/.style={
    code={
       \node[rectangle, draw=black, fill=white!5, very thick,
        minimum width=1.6cm, minimum height=0.6cm] (-body) at (0, 0) {\pgfkeysvalueof{/tikz/kraus/label}};
      \draw[thick] ([xshift=.5cm,yshift=0cm]-body.south)--++(0,-0.3cm) coordinate (-D) node[above]  {\pgfkeysvalueof{/tikz/kraus/D label}};
      \draw[thick] ([xshift=-.5cm,yshift=0cm]-body.south)--++(0,-0.3cm) coordinate (-C) node[above] {\pgfkeysvalueof{/tikz/kraus/C label}};
      \draw[thick] ([xshift=-.5cm,yshift=0cm]-body.north)--++(0,+0.3cm) coordinate (-A) node[below]  {\pgfkeysvalueof{/tikz/kraus/A label}};
    }
  },
    pics/kraust/.style={
    code={
      \node[rectangle, draw=black, fill=white!5, very thick,
        minimum width=1.6cm, minimum height=0.6cm] (-body) at (0, 0)  {\pgfkeysvalueof{/tikz/kraus/label}};
      \draw[thick] ([xshift=-.5cm,yshift=0cm]-body.south)--++(0,-0.3cm) coordinate (-D) node[above]  {\pgfkeysvalueof{/tikz/kraus/D label}};
      \draw[thick] ([xshift=+.5cm,yshift=0cm]-body.south)--++(0,-0.3cm) coordinate (-C) node[above] {\pgfkeysvalueof{/tikz/kraus/C label}};
      \draw[thick] ([xshift=.5cm,yshift=0cm]-body.north)--++(0,+0.3cm) coordinate (-A) node[below]  {\pgfkeysvalueof{/tikz/kraus/A label}};
    }
  },
}

\tikzset{
  pics/sigma/.style n args={1}{
    code={
      \node[
        rectangle,
        draw=black,
        fill=white!5,
        very thick,
        minimum size=.2cm
      ] (-body) at (0,0) {#1};
    }
  },
  pics/sigma/.default={$\sigma^{\frac12}$}
}

\NewDocumentCommand{\colE}{s m}{
  E\IfBooleanT{#1}{^{\dagger}}_{#2}
}

\NewDocumentCommand{\rowE}{s m}{
  \tilde{E}\IfBooleanT{#1}{^{\dagger}}_{#2}
}

\NewDocumentCommand{\petzChannel}{m}{
  \mathcal{R}^{\petz}_{#1}
}

\NewDocumentCommand{\petzKraus}{s m m}{
  R^{\petz\IfBooleanT{#1}{\dagger}}_{#2,#3}
}

\NewDocumentCommand{\colR}{s m}{
  R^{\petz\IfBooleanT{#1}{\dagger}}_{#2}
}

\NewDocumentCommand{\petzW}{s m}{
  W^{\petz\IfBooleanT{#1}{\dagger}}_{#2}
}

\NewDocumentCommand{\recoveryW}{s}{
  \IfBooleanTF{#1}{(W^{\vec\theta})^\dagger}{W^{\vec\theta}}
}

\NewDocumentCommand{\petz}{}{\mathrm{Petz}}
\NewDocumentCommand{\opt}{}{\mathrm{opt}}

\newcommand{\Fent}{F^{\mathrm{en}}}
\newcommand{\Std}{\operatorname{Std}}
\DeclareMathOperator{\sgn}{sign}

\begin{document}

\title{Petz Recovery through a Single Polar Transformation for Approximate Quantum Error Correction}
\author{Hanzhang Zhao}
\affiliation{Department of Physics \& Astronomy, Texas A\&M University, College Station, Texas 77843, USA}
\author{Shenglong Xu}
\email{slxu@tamu.edu}
\affiliation{Department of Physics \& Astronomy, Texas A\&M University, College Station, Texas 77843, USA}

\begin{abstract}
    The Petz recovery map provides a near optimal decoder for approximate quantum error correction, but its inverse square-root normalization complicates circuit implementation. We use the polar structure of the Petz recovery isometry to implement recovery through a single quantum singular value transformation (QSVT).
    We give explicit block-encoding circuits and an extended recovery isometry that yields approximate Petz recovery without postselection. We bound the diamond distance between the approximate and ideal Petz encoding-recovery channels in terms of a spectrum-weighted loss, which provides a channel specific objective for optimizing the QSVT angles. Uniform polynomial approximation yields an asymptotic error bound that decays exponentially with the number of QSVT iterations, at a rate controlled by the smallest nonzero scaled singular value. Applications to Haar-random encoding and the Hayden--Preskill protocol illustrate the tradeoff between circuit depth and recovery accuracy and place existing scrambling decoders within a common Petz recovery framework.
\end{abstract}

\maketitle

\tableofcontents

\section{Introduction}
Quantum error correction (QEC) protects quantum information by encoding it into a larger physical system and recovering it after noise acts~\cite{Shor1995Scheme,Knill1997Theory}. A successful recovery must preserve arbitrary logical superpositions and entanglement with external systems, rather than only restore classical populations. In exact QEC, the Knill--Laflamme conditions characterize when every state in a code can be recovered perfectly and allow errors to be diagnosed without revealing the encoded information~\cite{Knill1997Theory}.
% Beyond quantum memories and computation, the recovery of encoded information also provides a framework for studying information scrambling in quantum many-body systems, as exemplified by the Hayden--Preskill protocol~\cite{Hayden2007Black}. In each setting, the existence of a good recovery channel leaves a separate practical question: how can that channel be implemented?
Approximate quantum error correction~(AQEC) allows a controlled recovery error, relaxing the requirement of perfect reversibility and admitting a broader class of codes~\cite{Leung1997Approximate,Ng2010Simple,Beny2010General}. For a general noise channel $\E$, a closed-form expression for the optimal recovery channel $\R^{\opt}$ is generally unavailable, but the channel can be computed numerically using semidefinite programming~\cite{Fletcher2007Optimum}. The Petz recovery map provides a near-optimal analytical construction~\cite{Petz1986Sufficient,Petz1988Sufficiency}:
\begin{equation}\label{eq:petz_definition}
    \petzChannel{\sigma}(\cdot) = \sigma^{\frac{1}{2}}\mathcal{E}^\dagger[\mathcal{E}(\sigma)^{-\frac{1}{2}}(\cdot)\mathcal{E}(\sigma)^{-\frac{1}{2}}]\sigma^{\frac{1}{2}},
\end{equation}
which is determined by the noise channel $\E$ and a reference state $\sigma$. For a maximally mixed reference on the code space, it gives the transpose-channel recovery. 
The performance of the Petz recovery map is controlled by near-optimality guarantees for entanglement fidelity~\cite{Barnum2002Reversing,Ng2010Simple}. Conditions under which Petz recovery is optimal beyond exact QEC have also been characterized~\cite{Li2025Optimality}. Beyond AQEC, Petz recovery has applications in different fields. The pretty-good measurement~\cite{Hausladen1994Pretty}, which is a special case of Petz recovery, plays an important role in quantum state discrimination.
In the context of quantum many-body physics, Petz recovery and its rotated or twirled variants~\cite{Wilde2015Recoverability,Sutter2016Strengthened,Junge2018Universal} also appear in decoding scrambled information~\cite{Nakayama2023Petz}, controlling noisy dynamics~\cite{Kwon2022Reversing}, diagnosing quantum many-body phases through recovery fidelity~\cite{Hu2024Petz}, and establishing the stability of mixed-state phases through local recoverability and finite Markov length~\cite{Sang2025Stability}. In holography, both the twirled and ordinary Petz maps have been used for entanglement-wedge reconstruction in AdS/CFT~\cite{Cotler2019Entanglement,Chen2020Entanglement,Zhao2020Petz}.

The near-optimality of Petz recovery makes it a promising target for decoder construction. Recently, single-qubit recovery has been proposed for trapped ions~\cite{Png2025Petz} and demonstrated in NMR and photonic systems~\cite{Singh2026Realizing,Li2026Experimental}. A modified Petz recovery for a four-qubit code has also been demonstrated in logical-state lifetime experiments on IBM hardware~\cite{Biswas2026Universal}. For small systems, circuits for Petz recovery can be synthesized from classically computed recovery Kraus operators and unitary completion~\cite{Biswas2024NoiseAdapted,Biswas2026Universal}. When explicit classical computation becomes impractical, quantum algorithms offer an alternative. Their implementation must nevertheless handle the nonunitary factors $\E(\sigma)^{-1/2}$ and $\sigma^{1/2}$ and the adjoint map $\E^\dagger$. Quantum singular value transformation (QSVT) provides a unified framework for a broad class of quantum algorithms~\cite{Gilyen2019Quantum,Martyn2021Grand}. Gily\'en et al.~\cite{Gilyen2022Quantum} use QSVT to implement $\sigma^{1/2}$ and $\E(\sigma)^{-1/2}$ separately and combine these operations with amplitude amplification to approximate the full Petz channel. This approach requires access to a unitary dilation of the noisy channel and its inverse, which can be synthesized through Uhlmann transformations~\cite{Utsumi2025Quantum}. Alternatively, virtual recovery estimates Petz-recovered observables using only forward access to the noise channel, without requiring access to its dilation~\cite{Zhu2026Simulation}. Another approach is to simplify the recovery map itself. The Petz-lite construction~\cite{Nakayama2023Petz} drops the operator dependence of the inverse-square-root factor $\E(\sigma)^{-1/2}$ by replacing it with a scalar normalization. Related Petz-like decoders use phase-preserving fixed-point amplitude amplification to overcome low postselection success probabilities~\cite{Utsumi2026Explicit,Wang2025Relieving}. Under special conditions, the full Petz map itself admits a simpler realization. Work on tabletop reversibility~\cite{Aw2024Role,Song2025Exact} identifies conditions for exact implementation without separately implementing $\E(\sigma)^{-1/2}$ or using postselection.

The algorithm of Gily\'en et al.~\cite{Gilyen2022Quantum} implements the full Petz map for an arbitrary encoding channel $\E$ with guaranteed accuracy but requires separate implementations of its three component maps. It remains a question whether one can simplify this approach while retaining rigorous error guarantees. In this work, we develop a QSVT-based quantum algorithm that combines these components into a single singular-value \textit{sign} transformation. Our approach uses the representation of the Petz isometry as a polar factor~\cite{Ma2025Haar,Afham2026Projections,Elimelech2026Universal}. The polar structure of the pretty-good measurement, a special case of Petz recovery, was characterized in earlier work~\cite{Eldar2001Quantum} and subsequently used in a QSVT-based implementation~\cite{Quek2022Fast}.  Our formulation connects Petz recovery to simultaneous quantum searches with phase alignment across singular-vector sectors. In our formulation, the ideal limit of the Utsumi--Nakata decoders~\cite{Utsumi2026Explicit} can be identified with Petz recovery for a maximally mixed reference state $\sigma$. We construct an extended recovery isometry and give explicit block-encoding circuits that yield approximate recovery channels $\R^{\vec \theta}$ without postselection. These channels depend on the angle sequence $\vec\theta$ used in QSVT to approximate the sign function. We derive a spectrum-weighted bound on the diamond distance between the approximate and ideal encoding--recovery channels for a full-rank reference state, and use it as an objective for optimizing the QSVT angles:
\begin{equation}
    \begin{aligned}
    \frac{1}{2}\|\R^{\vec\theta}\circ\E-\petzChannel{\sigma}\circ\E\|_\diamond
    &\leq \sqrt{\|\sigma^{-1}\|_\infty\,\mathcal I^{\vec\theta}},\\
    \mathcal I^{\vec\theta}&=1-\left|\sum_i\lambda_i^2p^{\vec \theta}_k(\lambda_i/\alpha)\right|^2.
    \end{aligned}
\end{equation}
Here $p^{\vec{\theta}}_k$ is an odd complex QSVT polynomial of degree at most $2k+1$, determined by the angle sequence $\vec\theta$ at iteration count $k$. The $\lambda_i$ are the singular values of $\E(\sigma)^{1/2}$, and $\alpha$ is a constant determined by the block encoding used in QSVT. The infidelity $\mathcal I$ vanishes if $p(x) = \sgn(x)$. This bound provides an objective for optimizing the circuit angles $\vec \theta$ at finite $k$. Suitable angle sequences yield an analytical diamond-distance upper bound that decays as $\mathcal O(e^{-\lambda_{\min}k/\alpha})$ as $k\to\infty$, where $\lambda_{\min}$ is the smallest nonzero singular value~(Sec.~\ref{sec:convergence}). For a fixed reference state and spectrum, a target diamond-distance error $\varepsilon$ can therefore be achieved with an iteration count scaling as
\begin{equation}
    k=\mathcal O\!\left(\frac{\alpha}{\lambda_{\min}}\log\frac{1}{\varepsilon}\right),
    \qquad \varepsilon\to0.
\end{equation}

We then apply our construction to Haar-random encoding~\cite{Klesse2007Approximate,Hayden2008Decoupling,Ma2025Haar,Sommers2026Spectral} and the Hayden--Preskill protocol~\cite{Hayden2007Black}, illustrating the tradeoff between circuit depth and recovery accuracy. In the Hayden--Preskill setting, our construction unifies the deterministic and stochastic Yoshida--Kitaev decoders~\cite{Yoshida2017Efficient} and the Petz-lite decoder~\cite{Nakayama2023Petz} as special cases of a common Petz recovery framework, obtained through particular choices of iteration count and postselection.
\section{Setup and Notation}
\label{sec:setup}

\subsection{Encoding and Noise Channel}
% We consider a general quantum error-correction scenario in which a quantum state $\rho$ in the logical Hilbert space $\H_A$ is encoded into a physical Hilbert space $\H_D$ and subjected to noise. We denote the resulting composite quantum channel by $\mathcal{E}$, which maps the logical Hilbert space $\H_A$ to the physical Hilbert space $\H_D$. The channel $\mathcal{E}$ can be represented by a set of Kraus operators $\{E_i\}$, which satisfy the completeness relation $\sum_i E_i^\dagger E_i = I_A$ for any completely positive trace-preserving~(CPTP) channel. The action of the channel on a quantum state $\rho_A$ is given by
We describe encoding and noise by a completely positive, trace-preserving~(CPTP) channel $\mathcal{E}$ from the logical space $\H_A$ to the physical space $\H_D$. Its Kraus operators satisfy $\sum_i E_i^\dagger E_i=I_A$, and its action is
\begin{equation}
    \mathcal{E}(\rho_A) =  \sum_i E_i \rho E_i^\dagger.
\end{equation}

The Stinespring dilation theorem~\cite{Stinespring1955Positive,Kretschmann2008InformationDisturbance} states that any quantum channel can be realized by a unitary operator $U$ acting on \(\H_A\otimes\H_B\), with the auxiliary system $B$ initialized in $\ket{0}_B$, followed by a partial trace over the environment $\H_C$. Here \(\H_A\otimes \H_B \cong \H_C\otimes\H_D\).
In this formulation, the channel acts as
\begin{equation}
    \mathcal{E}(\rho) = \text{tr}_C\left(U\rho_A\otimes \ket{0}_B \bra{0}_B U^\dagger\right) =
    \begin{tikzpicture}[baseline={([yshift=-0.5ex]current bounding box.center)}]
        %nodes
        \pic[unitary/C label=$C$,unitary/D label=$D$, unitary/B label=$\ket{0}_B$]  (u1)        at(0,0)     {unitary};
        \pic[unitary/C label=$C$,unitary/D label=$D$, unitary/B label=$\bra{0}_B$]  (u2)        at(0,-2.5)     {unitaryd};
        \draw[dashed,thick] (u1-C) -- ++ (-1,0) |- (u2-C);
        \draw[thick] (u1-A)--(u2-A);
        \pic at ($(u1-A)!0.5!(u2-A)$) {sigma={$\rho_A$}};
        % %lines
    \end{tikzpicture},
\end{equation}
and the corresponding Kraus operator is
\begin{equation}
    E_i = \bra{i}_C\otimes I_D U I_A \otimes \ket{0}_B.
\end{equation}

The channel isometry $\colE{I}$ and its weighted version $\colE{\sigma}$ are defined by
\begin{equation}\label{eq:channel_isometry}
    \colE{I} = \sum_i \ket{i}_C \otimes E_i, \qquad \colE{\sigma} = \colE{I}\sqrt{\sigma},
\end{equation}
where $\sigma$ is a positive operator on $\H_A$. Both operators map $\H_A$ to $\H_C\otimes\H_D$. Kraus completeness gives
\begin{equation}
    \colE*{I} \colE{I} = I_A, \qquad \colE*{\sigma}\colE{\sigma}=\sigma.
\end{equation}
Thus $\colE{I}$ is an isometry. The encoded state can be expressed as
\begin{equation}
    \mathcal{E}(\rho) = \mathrm{tr}_C \left(\colE{\rho} \colE*{\rho}\right).
\end{equation}

\subsection{Decoding Channel}
In the decoding stage, a decoding channel $\mathcal{R}$ from $\mathcal{H}_D$ to $\mathcal{H}_A$ is applied to $\mathcal{E}(\rho)$, aiming to reverse $\mathcal{E}$ and recover $\rho$. Like $\mathcal{E}$, the channel $\mathcal{R}$ is characterized by a set of Kraus operators $\{R_i\}$ and can be realized by the isometry
\[
    R=\sum_i \ket{i}_C\otimes R_i,
\]
acting on $\mathcal{H}_D$, followed by tracing out the recovery environment $C$.

% In the case of perfect error correction, the composite channel $\mathcal{R} \circ \mathcal{E}$ returns the input state $\rho$ for every logical input. Perfect error correction is characterized by the Knill--Laflamme~(KL) condition~\cite{Knill1997Theory}:
% \begin{equation}\label{eq:knill_laflamme}
%     E_i^\dagger E_j = c_{ij} I_A.
% \end{equation}
% Here $c_{ij}$ are the entries of a positive semidefinite matrix of unit trace, independent of the logical input state.

The performance of the decoding channel can be characterized by the entanglement fidelity, which measures how well the channel preserves entanglement with a reference system $A'$ when $A'A$ is initially in a joint pure entangled state $\ket{\tau}$. The entanglement fidelity is defined as
\begin{equation}
\Fent_\tau = \bra{\tau}_{A'A}\left[\R\circ\E(\ket{\tau}_{A'A}\bra{\tau}_{A'A})\right]\ket{\tau}_{A'A},
\end{equation}
where the composite channel $\R\circ\E$ acts on the logical system $A$ and the reference system $A'$ is left untouched.
A diagrammatic representation is given in Eq.~\eqref{eq:ent_fidelity_diagram} of Appendix~\ref{sec:near_optimality}.

The entanglement fidelity for the maximally entangled EPR state, denoted simply by $\Fent$, is particularly important because it is directly related to the average fidelity over all possible logical states drawn from the Haar-random ensemble~\cite{Nielsen2002Simple}:
\begin{equation}
    F_{\text{avg}} = \frac{d_A \Fent + 1}{d_A + 1}.
\end{equation}
When the recovered output state is independent of the logical input, the entanglement fidelity is $\Fent = 1/d_A^2$, while for exact QEC, the optimal entanglement fidelity is $\Fent = 1$. For approximate QEC, the optimal decoding channel $\mathcal{R}^{\opt}$ achieves the highest possible entanglement fidelity given the noise channel $\mathcal{E}$. While a closed-form expression for the optimal decoding channel is generally unavailable, the Petz recovery map provides a near-optimal closed-form decoder.

\section{Petz Recovery and Its Circuit Representation}
\label{sec:maps}
\subsection{Kraus Representation and Recovery Guarantees}
The Petz map in Eq.~\eqref{eq:petz_definition} uses the adjoint map $\mathcal{E}^\dagger$ and the inverse square root $\E(\sigma)^{-\frac{1}{2}}$, defined on the support of $\E(\sigma)$ and set to zero on its kernel. Its Kraus operators are
\begin{equation}\label{eq:petz_kraus}
    \petzKraus{\sigma}{i} = \sigma^{\frac{1}{2}}E^\dagger_i\mathcal{E}(\sigma)^{-\frac{1}{2}},
\end{equation}
where $E_i$ are the Kraus operators of the channel $\mathcal{E}$. Checking the completeness relation of the Petz map yields
\begin{equation}
    \begin{aligned}
    \sum_i \petzKraus*{\sigma}{i} \petzKraus{\sigma}{i}
    &= \mathcal{E}(\sigma)^{-\frac{1}{2}}\mathcal{E}(\sigma)\mathcal{E}(\sigma)^{-\frac{1}{2}} \\
    &= \Pi_{\operatorname{supp}\mathcal{E}(\sigma)}.
    \end{aligned}
\end{equation}
Therefore, in general, the Petz map is positive but does not preserve the trace, and it becomes CPTP only when the state $\E(\sigma)$ is full rank in $D$. In Appendix~\ref{sec:completeness}, we show that the composite channel $\petzChannel{\sigma}\circ\E$ is a trace-preserving channel when the reference state $\sigma$ in the logical space $A$ is full rank. Throughout this work, we only assume that the reference state $\sigma$ in the logical space is full rank.

The Petz map achieves perfect recovery whenever exact error correction is possible and satisfies a near-optimality guarantee for approximate error correction. Its entanglement fidelity is bounded below by the square of the entanglement fidelity of the optimal recovery channel~\cite{Barnum2002Reversing}:
\begin{equation}
    \Fent_\sigma(\mathcal{R}^{\opt}\circ\mathcal{E}) \geq \Fent_\sigma(\petzChannel{\sigma}\circ\mathcal{E}) \geq \Fent_\sigma(\mathcal{R}^{\opt}\circ\mathcal{E})^2,
\end{equation}
where the entanglement fidelity is evaluated using a purification of the reference state $\sigma$. A simple proof based on the Cauchy--Schwarz inequality is presented in Appendix~\ref{sec:near_optimality}.

\subsection{Circuit Representation}
In diagrammatic tensor-network notation, the Petz map acting on the output of the encoding channel takes the form
\begin{equation}\label{eq:petz}
    \begin{tikzpicture}[baseline={([yshift=-0.5ex]current bounding box.center)}]
    \pic[unitary/B label=$\ket{0}_B$, unitary/A label=$\ket{\psi}_A$]  (u1)        at(0,0)     {unitary}; 
    \pic (u2)        at(0, 2)     {unitaryd};
    \draw[thick] (u2-A)--++(0,0.7) node[above] {$A$};
    \draw[thick] (u2-B)--++(0,0.7) node[above] {$\bra{0}_B$};
    \draw[thick] (u1-D) -- (u2-D);
    \pic at ([yshift=0.25cm]u2-A) {sigma={$\sigma^{\frac{1}{2}}$}}; 
    \pic at ($(u1-D)!0.5!(u2-D)$) {sigma={$\E(\sigma)^{-\frac{1}{2}}$}};  
    \end{tikzpicture}=
    \begin{tikzpicture}[baseline={([yshift=-0.5ex]current bounding box.center)}]
    \pic[unitary/B label=$\ket{0}_B$, unitary/A label=$\ket{\psi}_A$]  (u1)        at(0,0)     {unitary}; 
    \pic (u2)        at(0, 2)     {unitaryd};
    \draw[thick] (u2-A)--++(0,0.7) coordinate (A) node[above]{$A$};
    \pic at ([yshift=0.25cm]u2-A) {sigma={$\sigma^{\frac{1}{2}}$}}; 
    \draw[thick] (u2-B)--++(0,0.7) node[above] {$\bra{0}_B$};
    \draw[thick] (u1-D) -- (u2-D);
    \pic at ([xshift=0.8cm]$(u1-D)!0.5!(u2-D)$) {sigma={$\sqrt{d_C}\E(\sigma)^{-\frac{1}{2}}$}};
    \draw[thick] (u1-C) -- ++ (-1,0.5)coordinate (turn)--  (turn |- A) node[above] {$C$};
    \draw[thick] (u2-C) -- ++ (-0.7,0) coordinate (turn)--  (turn |- A) node[above] {$C'$};
    \filldraw ([xshift=-0.35cm]u2-C) circle (2pt);
    \node at (u2-C) [xshift=0.35cm] {$C''$};
    \end{tikzpicture}.
\end{equation}
The primed subsystem labels $C'$ and $C''$ indicate the ancillary systems with dimension $d_C$, and $\bullet$ represents a scalar factor $1/\sqrt{d_C}$ required for a normalized EPR pair between $C'$ and $C''$.

Tracing out $CC'$ yields the recovered state $\petzChannel{\sigma} \circ \E(\ket{\psi_A}\bra{\psi_A})$. The diagram represents the following sequence of operations:
\begin{enumerate}
    \item Prepare an EPR pair in subsystems $C'C''$, retaining $C'$ as the recovery environment. The normalization factor $\sqrt {d_C}$ is included in the operator $\E(\sigma)^{-\frac{1}{2}}$.
    \item Apply the nonunitary operator $\sqrt{d_C}\mathcal{E}(\sigma)^{-\frac{1}{2}}$ to subsystem $D$.
    \item Apply $U^\dagger$ to $C''D$, and postselect subsystem $B$ onto $\ket{0}_B$.
    \item Apply the operator $\sigma^{\frac{1}{2}}$ to subsystem $A$.
\end{enumerate}

Implementing each step above individually can be challenging. The operators $\mathcal{E}(\sigma)^{-\frac{1}{2}}$ in step~2 and $\sigma^{\frac{1}{2}}$ in step~4 are generally nonunitary and require synthesis. The postselection in step~3 is probabilistic and may have an exponentially small success probability.

% several of its steps are individually non-trace-preserving. This leads to two practical obstacles: Implementing $\mathcal{E}^{-\frac{1}{2}}(\sigma)$ generally requires matrix inverses and square roots, which are costly to compute and non-unitary and the success probability of Step~3 can be exponentially small, necessitating additional techniques for amplitude boosting.

Gily\'en et al.~\cite{Gilyen2022Quantum} implement the inverse-square-root factor in step 2 and the square-root factor in step 4 using QSVT, and use oblivious amplitude amplification to overcome the postselection in these steps as well as step 3 in order to obtain a deterministic approximation to the full Petz channel. Thus, the postselection in the factorized circuit is removed by coherent amplification. For a maximally mixed reference state $\sigma=I_A/d_A$, step~4 contributes only a scalar factor. The Petz-lite construction~\cite{Nakayama2023Petz} additionally replaces the operator-dependent normalization in step~2 with a scalar, leaving a recovery map proportional to $\E^\dagger$, implemented by postselecting $B$ onto $\ket{0}_B$ in step~3. Related Petz-like decoders replace this postselection with phase-preserving fixed-point amplitude amplification~\cite{Utsumi2026Explicit,Wang2025Relieving}. 

Our construction instead combines the three components into a single singular-value sign transformation. The next section introduces the operator underlying this polar representation and uses it to construct an extended recovery isometry and explicit circuits for approximate Petz recovery without postselection. 

\section{Polar-Based Petz Recovery}
\label{sec:polar_recovery}

\subsection{Petz Recovery as a Singular-Value Sign Transformation}
We define the Petz isometry by stacking the Kraus operators in Eq.~\eqref{eq:petz_kraus}:
\begin{equation}
    \colR{\sigma} = \sum_i \ket{i}_C \otimes \petzKraus{\sigma}{i}=\sum_i \ket{i}_C \otimes\sigma^{\frac{1}{2}}E^\dagger_i\mathcal{E}(\sigma)^{-\frac{1}{2}}.
\end{equation}
To expose its polar structure, we introduce
\begin{equation}\label{eq:weighted_operator}
    \rowE{\sigma}=\sum_i\bra{i}_C\otimes E_i\sqrt{\sigma},
\end{equation}
which maps $\H_C\otimes\H_A$ to $\H_D$. This definition applies to any positive weight, including $I_A$, and gives
\begin{equation}
    \E(\rho)=\rowE{\rho}\rowE*{\rho}.
\end{equation}
% In particular, the KL condition in Eq.~\eqref{eq:knill_laflamme} can be compactly restated as
% \begin{equation}
%     \rowE*{I}\rowE{I}=\rho_C\otimes I_A,
% \end{equation}
% where $\rho_C=\sum_{ij}c_{ij}\ket{i}_C\bra{j}_C$ is a density matrix on the Kraus-index space.
% Crucially, using $\E(\sigma)=\rowE{\sigma}\rowE*{\sigma}$, 
Thus we can express the Petz isometry as
\begin{equation}
    \colR{\sigma} = \rowE*{\sigma} \left(\rowE{\sigma} \rowE*{\sigma}\right)^{-\frac{1}{2}} = \sgn(\rowE*{\sigma}),
\end{equation}
where the sign function is applied to the singular values, with $\sgn(0)=0$. Thus the inverse-square-root normalization is absorbed into the polar factor of $\rowE*{\sigma}$.

Consider the compact singular value decomposition of $\rowE*{\sigma}$,
\begin{equation}
    \rowE*{\sigma} = \sum_ i \lambda_i \ket{E_i^{CA}}\bra{E_i^{D}},
\end{equation}
where $\lambda_i>0$, and $\ket{E_i^{CA}}$ and $\ket{E_i^{D}}$ are the left and right singular vectors in $\H_C\otimes\H_A$ and $\H_D$, respectively. The singular values obey $\sum_i\lambda_i^2=1$ because $\rowE{\sigma}\rowE*{\sigma}=\E(\sigma)$ is a density matrix. The Petz isometry can be expressed as
\begin{equation}
    \colR{\sigma} =\sum_i \ket{E_i^{CA}} \bra{E_i^{D}},
\end{equation}
which maps $\ket{E_i^{D}}$ to $\ket{E_i^{CA}}$ and annihilates states orthogonal to $\operatorname{supp}\E(\sigma)$.

\subsection{Extended Petz Isometry}
Since $\colR*{\sigma}\colR{\sigma}=\Pi_{\operatorname{supp}\E(\sigma)}$, the Petz isometry preserves norms only on the encoded support. To complete it to an isometry on all of $\H_D$, introduce normalized auxiliary states $\ket{s}_S$ and $\ket{t}_T$ in an extended space
\begin{equation}
    \H_{\mathrm{ext}} \cong \H_S \otimes \H_D \cong \H_T \otimes \H_C \otimes \H_A,
\end{equation}
of dimension $d_{\mathrm{ext}}$. With sufficiently large auxiliary spaces, choose a unitary $\petzW{\sigma}$ satisfying
\begin{equation}\label{eq:W_Petz}
    \bra{t}\petzW{\sigma}\ket{s}=\colR{\sigma}=\sum_i \ket{E_i^{CA}}\bra{E_i^{D}}.
\end{equation}
Together with unitarity, this condition maps $\ket{s}\otimes\ket{E_i^{D}}$ to $\ket{t}\otimes\ket{E_i^{CA}}$. From this, it follows that
\begin{equation}\label{eq:extended_petz_isometry}
    \begin{aligned}
    \petzW{\sigma}\ket{s} &= \ket{t}\otimes\colR{\sigma} \\
    &\quad +(I_T-\ket{t}\bra{t})W^\perp\ket{s}
    (I_D-\sum_i \ket{E_i^{D}}\bra{E_i^{D}}).
    \end{aligned}
\end{equation}
Here $W^\perp$ is any unitary on $\H_{\mathrm{ext}}$ satisfying
\[
    \bra{t}W^\perp\ket{s}
    \left(I_D-\sum_i\ket{E_i^D}\bra{E_i^D}\right)=0,
\]
so inputs orthogonal to $\operatorname{supp}\E(\sigma)$ are mapped to outputs orthogonal to the auxiliary state $\ket{t}$.

We call $\petzW{\sigma}\ket{s}$ the extended Petz isometry. Postselecting $T$ onto $\ket{t}$ and discarding $C$ implements the original Petz map in Eq.~\eqref{eq:petz_definition} for any input. The success probability equals the input's weight on $\operatorname{supp}\E(\sigma)$. Discarding both $T$ and $C$ instead defines a trace-preserving extension of the Petz map. Since $\sigma$ is full rank, every encoded state lies in this support, so postselection succeeds with certainty and can be omitted:
\begin{equation}\label{eq:extended_petz}
    \begin{aligned}
    &\tr_{TC}\left[\petzW{\sigma}
    \left(\ket{s}\bra{s}\otimes\E(\rho)\right)\petzW*{\sigma}\right]  =\petzChannel{\sigma}\circ\E(\rho).
    \end{aligned}
\end{equation}
Full rank of $\sigma$ is sufficient for this agreement on encoded states; $\E(\sigma)$ need not be full rank.

To construct this isometry $\petzW{\sigma}\ket{s}$, we start from a block encoding of $\rowE*{\sigma}$,
\begin{equation}
    \rowE*{\sigma}=\alpha\bra{t}W\ket{s},
\end{equation}
where $W$ is unitary and $\alpha>0$. The block encoding specifies the auxiliary registers and states; explicit choices are given in Sec.~\ref{sec:block_encoding_circuits}. QSVT then approximates the desired singular-value sign transformation,
\begin{equation}\label{eq:W_blockencoding}
   \bra{t}W\ket{s}\xrightarrow{\mathrm{QSVT}} \bra{t}\petzW{\sigma}\ket{s}
    =\sgn(\bra{t}W\ket{s})=\colR{\sigma}.
\end{equation}

This construction also has a quantum-search interpretation. After applying $W$, define the initial and target search states in $\H_T \otimes \H_C \otimes \H_A$ as follows:
\begin{equation}
    \begin{aligned}
    \ket{s_i}=W\bigl(\ket{s}\otimes\ket{E_i^{D}}\bigr),\ \ket{t_i}=\ket{t}\otimes\ket{E_i^{CA}}.
    \end{aligned}
\end{equation}
From the SVD of $\rowE*{\sigma}$, their overlaps are
\begin{equation}
    \braket{t_i|s_j}=x_i\delta_{ij},\qquad x_i=\lambda_i/\alpha.
\end{equation}
Eq.~\eqref{eq:extended_petz_isometry} specifies the desired search after this preparation:
\begin{equation}
    \ket{s_i}\xrightarrow{\text{quantum search}} \petzW{\sigma}W^\dagger\ket{s_i}=\ket{t_i}.
\end{equation}
The pairs $\{\ket{s_i},\ket{t_i}\}$ span mutually orthogonal subspaces of dimension at most two. Recovery therefore amounts to simultaneous searches in these subspaces with different overlaps $x_i$, while preserving relative phases between sectors. Standard Grover iterations generally overshoot at different iteration counts for different overlaps. We instead use fixed-point quantum search~\cite{Grover2005FixedPoint,Yoder2014FixedPoint} with the additional requirement of phase alignment. 

This also explains the connection to the Petz-lite decoder~\cite{Nakayama2023Petz} for $\sigma=I_A/d_A$. With an appropriate block encoding, applying $W$ and postselecting the target register $T$ onto $\ket{t}$ gives a recovery map proportional to $\E^\dagger$. The stochastic Yoshida--Kitaev decoder~\cite{Yoshida2017Efficient} has a similar postselected realization (Sec.~\ref{sec:existing_decoders}). Utsumi and Nakata~\cite{Utsumi2026Explicit} construct decoders by replacing postselection with fixed-point amplitude amplification. Their decoupling-based analysis emphasizes that successful recovery requires phase alignment across singular-vector sectors. Here, this requirement follows directly from the polar representation of the Petz isometry: the ideal search maps each $\ket{s_i}$ to $\ket{t_i}$ with the same phase.

\subsection{QSVT Construction and Error Bound}
To construct the QSVT sequence, we introduce two unitary rotations on the extended Hilbert space:
\begin{equation}
    \begin{aligned}
    &Z_s(\theta) = W\left[\exp\left(i\theta\ket{s}\bra{s}\right)\otimes I_D\right]W^\dagger,\\
    &Z_t(\theta) = \exp\left(i\theta\ket{t}\bra{t}\right)\otimes I_{CA}.
    \end{aligned}
\end{equation}
Within each search subspace, $Z_s$ and $Z_t$ apply phase shifts to $\ket{s_i}$ and $\ket{t_i}$, respectively. We first apply $W$ to prepare the search states and then alternate these rotations:
\begin{equation}
    \begin{aligned}
    \recoveryW & = Z_s(\theta_{2k})Z_t(\theta_{2k-1})\cdots Z_s(\theta_2)Z_t(\theta_1)W\\
    &= \left\{\prod_{i=1}^k\left[Z_s(\theta_{2i})Z_t(\theta_{2i-1})\right]\right\}W.
    \end{aligned}
\end{equation}
Here $Z_t$ and $Z_s$ play the roles of a generalized Grover oracle and diffusion operator, respectively. The sequence of rotations is parameterized by a set of angles $\vec{\theta} = (\theta_1, \theta_2, \cdots, \theta_{2k})$. In standard Grover search, the rotation angles are fixed to $\theta_i = \pi$ for all $i$, while in fixed-point Grover search, the rotation angles are chosen to be different for each iteration. In QSVT, the rotation angles are chosen to approximate a polynomial transformation of the singular values of $\bra{t} W \ket{s}$.

By expanding the rotations in their defining projectors, one can show that $\recoveryW\ket{s}$ decomposes into two orthogonal terms:
\begin{equation} \label{eq:W_qsvt}
    \recoveryW\ket{s} = \ket{t}P+(I_T-\ket{t}\bra{t})W\ket{s}Q,
\end{equation}
where the operators $P$ and $Q$ are defined by
\begin{equation}
    \begin{aligned}
    P &= \sum_i p(x_i)\ket{E_i^{CA}}\bra{E_i^{D}},\\
    Q &= \sum_i q(x_i)\ket{E_i^{D}}\bra{E_i^{D}}\\
      &\quad + q(0)\left(I_D-\sum_i\ket{E_i^{D}}\bra{E_i^{D}}\right).
    \end{aligned}
\end{equation}
Here $p$ and $q$ are scalar odd and even polynomials of degree at most $2k+1$ and $2k$, respectively, determined by the rotation angles $\theta_i$.
%The operator $P$ maps right singular vectors to left singular vectors, while $Q$ acts diagonally in the input singular-vector basis and as $q(0)$ on states orthogonal to $\operatorname{supp}\E(\sigma)$.
The isometry condition $\bra{s}\recoveryW* \recoveryW\ket{s}=I_D$ then implies the constraint
\begin{equation}
    |p(x)|^2 + (1-x^2)|q(x)|^2 = 1,
\end{equation}
for all singular values of $\bra{t}W\ket{s}$. For instance, at $k=0$, $p(x) = x$ and $q(x) = 1$.

From a QSVT perspective, $\recoveryW$ implements a singular value transformation:
\begin{equation}
   x_i \xrightarrow{\mathrm{QSVT}} p(x_i).
\end{equation}
From a quantum search perspective, $\recoveryW W^\dagger$ implements simultaneous searches in the subspaces spanned by $\{\ket{s_i},\ket{t_i}\}$:
\begin{equation}
    \ket{s_i}\xrightarrow{\text{Quantum Search}} p(x_i)\ket{t_i} + q(x_i)\sqrt{1-x_i^2}\ket{t_i^\perp},
\end{equation}
where $|p(x_i)|^2$ is the fidelity between the final state and the target state, and $\ket{t_i^\perp}$ is the normalized state orthogonal to the target state:
\begin{equation}
    \ket{t_i^\perp} = \frac{\ket{s_i}-x_i\ket{t_i}}{\sqrt{1-x_i^2}}.
\end{equation}
When $x_i=1$, this component vanishes and no perpendicular state is needed.

By comparing with Eqs.~\eqref{eq:W_Petz} and \eqref{eq:W_blockencoding}, we see that the Petz map corresponds to
\begin{equation}\label{eq:PQ_Petz}
    \begin{aligned}
    p^{(\mathrm{Petz})}(x) = \sgn(x)
    \end{aligned}
\end{equation}
apart from overall phase factors. Successful fixed-point search only requires $|p(x_i)|^2 = 1$, allowing for an inhomogeneous phase factor for each singular value. On the other hand, the Petz map requires $p(x_i) = 1$ up to a global phase for all nonzero singular values of $\bra{t}W\ket{s}$, which is a stronger requirement.

The ideal transformations can be approximated by a finite-degree polynomial in the singular values, which allows for practical implementation. Importantly, unlike general QSVT, postselecting the $\ket{t}$ state is optional because, when $p$ is close to the sign function, the error component is automatically suppressed. As a result, we can use the full channel to approximate the Petz map without postselection. We denote the approximate Petz recovery channel by
\begin{equation}
    \R^{\vec\theta}(\rho) =\mathrm{tr}_{T\otimes C}\recoveryW \ket{s}\bra{s}\otimes \rho  \recoveryW*.
\end{equation}
Appendix~\ref{sec:diamond_distance} bounds the diamond distance between the composite encoding--recovery channels as
\begin{equation}\label{eq:diamond_norm_bound}
    \Delta(\vec \theta) = \frac{1}{2}\|\R^{\vec \theta} \circ \E - \petzChannel{\sigma} \circ \E\|_\diamond \leq \sqrt{\|\sigma^{-1}\|_{\infty} \mathcal{I}^{\vec{\theta}}},
\end{equation}
where 
\begin{equation}
    \mathcal{I}^{\vec{\theta}} = 1 - \left|\sum_i \lambda_i^2 p(x_i)\right|^2.
\end{equation}
The quantity $|\sum_i\lambda_i^2p(x_i)|^2$ is the fidelity between the pure outputs of our approximate and ideal Petz encoding--recovery isometries for a normalized purification $\ket{\sigma}_{AA'}$ of $\sigma$, with $A'$ untouched and all environment and auxiliary registers retained (Appendix~\ref{sec:diamond_distance}). Thus $\mathcal I^{\vec\theta}$ is the infidelity between these two pure output states.

Recall that $\sum_i \lambda_i^2 = 1$. Thus the right-hand side is small when $p(x_i)$ is close to 1 up to a global phase factor for all nonzero singular values $x_i$, in which case the approximate and ideal Petz encoding--recovery channels are close in diamond distance.

We compare the composite channels because the original Petz map is trace preserving only on $\operatorname{supp}\E(\sigma)$. With the inverse set to zero outside this support, it is generally trace-nonincreasing rather than CPTP on all of $\H_D$. For full-rank $\sigma$, every encoded state lies within this support, so both composite channels are CPTP and their comparison avoids an arbitrary choice of trace-preserving extension.

% Now we present our polar-based Petz recovery algorithm in Algorithm~\ref{alg:petz}. The algorithm takes as input the encoded state $\E(\rho)$ and the encoded reference state $\E(\sigma)$, and outputs a state in subsystem $A$ that is approximately the Petz recovered state $\petzChannel{\sigma} \circ \E(\rho)$ by applying the unitary operator $\recoveryW$ to the composite state $\ket{s}\bra{s}\otimes \E(\rho)$ in the extended Hilbert space $\H_{\mathrm{ext}}$.
Algorithm~\ref{alg:petz} summarizes the polar-based Petz recovery procedure. Without the optional postselection step, the algorithm implements $\R^{\vec\theta}\circ\E$, and the diamond-distance bound in Eq.~\eqref{eq:diamond_norm_bound} applies. With postselection, it instead produces the normalized successful branch associated with $P$; the bound above does not apply to this conditional output.
Explicit block-encoding constructions and their access requirements are given in Sec.~\ref{sec:block_encoding_circuits}. The quantum sequence uses $k+1$ applications of $W$ and $k$ of $W^\dagger$, together with $2k$ auxiliary-state phase rotations and the initial preparation of $\ket{s}$. 

\begin{algorithm}
\caption{Petz Recovery via a Single Polar Transformation}
\label{alg:petz}
\begin{algorithmic}[1]
\Require Encoded state $\E(\rho)$; a block encoding $\alpha\bra{t}W\ket{s}=\rowE*{\sigma}$ for a full-rank reference state $\sigma$, with known $\alpha>0$.
\Require Coherent access to $W$, $W^\dagger$, and preparation unitaries for $\ket{s}$ and $\ket{t}$ and their inverses.
\Require Classical knowledge of the nonzero singular values $\{\lambda_i\}$, or a certified interval $[a,b]\subseteq(0,1]$ containing all $x_i=\lambda_i/\alpha$.
\State Choose the number of iterations $k\geq0$.
\State Compute the angles $\vec\theta$ classically using spectrum-weighted or interval-based optimization~(Sec.~\ref{sec:angle_finding}).
\State Prepare the initial state $\ket{s}$ in the auxiliary Hilbert space $\H_S$.
\State Apply $W$ to $\ket{s}\bra{s}\otimes\E(\rho)$ and denote the resulting state in $\H_{\mathrm{ext}}$ as $\rho_{ext}^{(0)}$.
\For{$i=1,2,\cdots, k$}
    \State Apply the unitary operator $Z_s(\theta_{2i})Z_t(\theta_{2i-1})$ to the state $\rho_{ext}^{(i-1)}$ to obtain the state $\rho_{ext}^{(i)}$.
\EndFor
\State \textbf{Optional:} Postselect register $T$ onto $\ket{t}$, retaining only successful runs.
\State Discard the auxiliary registers $T$ and $C$.
\Ensure The recovered state on $A$, conditioned on success if postselection is used.
\end{algorithmic}

\begin{tikzpicture}[
    gate/.style={
        rectangle,
        draw=black,
        fill=white!5,
        very thick,
        minimum width=1.6cm,
        minimum height=0.5cm
    },
    small gate/.style={
        rectangle,
        draw=black,
        fill=white,
        thick,
        minimum width=0.8cm,
        minimum height=0.5cm
    }
]

% Main circuit nodes
\pic[
    unitary/A label=$\ket{\psi}_A$,
    unitary/B label=$\ket{0}_B$
] (u1) at (0,0) {unitary};

\node[gate] (w)    at (1,1) {$W$};
\node[gate] (wone) at (1,2) {$W_1$};
\node[gate] (wi)   at (1,3) {$W_i$};
\node[gate] (wk)   at (1,4) {$W_k$};

\node[anchor=north] (s) at ([xshift=1cm]u1-B) {$\ket{s}$};
\draw[thick] (s) -- (s |- w.south);
\draw[thick] (u1-D) -- (u1-D |- w.south);
\draw[thick] (w.north) -- (wone.south);
\node at ($(wone.south)!0.5!(wi.north)$) {$\cdots$};
\node at ($(wi.south)!0.5!(wk.north)$) {$\cdots$};
\draw[thick] ([xshift=-0.5cm]wk.north) -- ++(0,0.25cm) node[above] {$C$};
\draw[thick] (wk.north) -- ++(0,0.25cm) node[above] {$A$};
\draw[thick] ([xshift=0.5cm]wk.north) -- ++(0,0.25cm) node[above] {$T$};
\draw[thick] (u1-C) -- ([yshift=0.25cm]u1-C |- wk.north) node[above] {$C$};

% Arrow from W_i to the expanded diagram
\draw[->, thick]
    (wi.east) -- ++(1.0,0)
    coordinate (expand-start);

% Four connected gates
\node[gate] (g1) at ([xshift=1.2cm, yshift=1cm]expand-start){$W$};
\node[gate, below=0.45cm of g1] (g2) {$e^{i\theta_{2i}\ket{s}\bra{s}}$};
\node[gate, below=0.45cm of g2] (g3) {$W^\dagger$};
\node[gate, below=0.45cm of g3] (g4) {$Z_t(\theta_{2i-1})$};
\draw[thick] ([yshift=0.25cm]g1.north)--(g1.north) (g1.south) -- (g2.north) (g2.south) -- (g3.north) (g3.south) -- (g4.north) (g4.south) -- ++(0,-0.25cm);

% Dashed box around the expanded gates
\node[
    draw,
    dashed,
    thick,
    rounded corners,
    inner sep=7pt,
    fit=(g1)(g2)(g3)(g4),
    label=above:expanded $W_i$
] (expanded-box) {};

\end{tikzpicture}
\end{algorithm}

\section{Angle Optimization and Circuit Implementation}
\label{sec:optimization_implementation}
The two remaining ingredients for implementing our algorithm are the choice of rotation angles $\theta_i$ and the block encoding of the operator $\rowE*{\sigma}$. We first discuss angle optimization and convergence, then give explicit block-encoding circuits.

\subsection{Spectrum-Weighted and Interval-Based Optimization}\label{sec:angle_finding}
Because $Z_s$ and $Z_t$ preserve each subspace spanned by $\ket{s_i}$ and $\ket{t_i}$, we can evaluate $p(x)$ using the single-qubit quantum signal processing (QSP) representation~\cite{Low2017Optimal}:
\begin{equation}
    p(x)\ket{0} + q(x)\sqrt{1-x^2}\ket{1} = w(x)\prod_{i=1}^{2k}\bigl[z(\theta_i)w(x)\bigr]\ket{0},
\end{equation}
where \(w(x)\) and \(z(\theta_i)\) are the 2-by-2 unitary matrices
\begin{equation}
    w(x) = \begin{pmatrix}
        x & -\sqrt{1-x^2}\\
        \sqrt{1-x^2} & x
        \end{pmatrix},\quad
    z(\theta_i) = \begin{pmatrix}
        e^{i\theta_i} & 0\\
        0 & -1
    \end{pmatrix}.
\end{equation}
Here $\ket{0}$ and $\ket{1}$ represent $\ket{t_i}$ and $\ket{t_i^\perp}$, respectively. This representation allows classical angle optimization using only 2-by-2 matrix products, without simulating $\recoveryW$ on the full extended space.

The bound on the diamond distance in Eq.~\eqref{eq:diamond_norm_bound} provides a practical criterion for optimization, as it only depends on the sum of the polynomial $p(x)$ weighted by the spectrum $\lambda$. We optimize over the odd polynomials $p(x)$ of degree at most $2k+1$ that are realizable by the QSVT sequence. In particular, $|p(x)|\leq 1$ for $x\in[-1,1]$. The spectrum-weighted optimization problem is:
\begin{equation}\label{eq:optimal_polynomial}
    \text{weighted:}\quad 1-\delta^{\mathrm w}(k) = \max\limits_{p} \left | \sum_i \lambda^2_i p(x_i) \right |.
\end{equation}
The superscript $\mathrm w$ denotes spectrum-weighted optimization. The infidelity in the bound is
\begin{equation}
    \mathcal I_k^{\mathrm w}=1-\left(1-\delta^{\mathrm w}(k)\right)^2.
\end{equation}

\begin{figure}[tbp]
    \centering
    \includegraphics[width=\columnwidth]{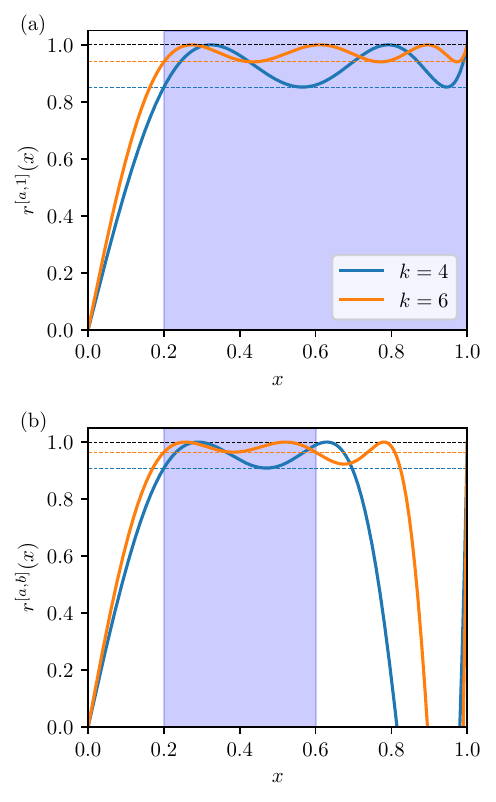}
    \caption{Real odd polynomial approximations $r(x)$ to the sign function for $k=4$ (blue) and $k=6$ (orange) QSVT iterations. These polynomials are realized as $r(x)=\operatorname{Re}[e^{-i\phi}p(x)]$ after a common phase choice. (a) Optimization on $[0.2,1]$. (b) Optimization on $[0.2,0.6]$. The shaded regions mark the optimization intervals. The black dashed lines mark the target value $1$, and the colored dashed lines mark the minimum polynomial values on the respective intervals. Increasing $k$ or restricting the optimization interval improves the approximation.}
    \label{fig:optimal_poly}
\end{figure} 

Spectrum-weighted optimization uses the individual singular values $x_i$, which may be unavailable for large systems. If only an interval $[a,b]$ containing all nonzero $x_i$ is known, we can instead optimize a real odd polynomial $r(x)$ of degree at most $2k+1$, subject to $|r(x)|\leq1$ for $x\in[-1,1]$:
\begin{equation}\label{eq:optimal_polynominal_minmax}
    \text{interval:}\quad 1-\delta^{[a,b]}(k)
    =\max_r\min_{x\in[a,b]}r(x).
\end{equation}
By the QSP polynomial-completion theorem~\cite{Gilyen2019Quantum}, any such $r$ can be realized as $r(x)=\operatorname{Re}[e^{-i\phi}p(x)]$, where $p$ is the complex polynomial produced by the QSVT sequence. The phase $\phi$ is common to all $x$: multiplying $\recoveryW$ by $e^{-i\phi}$ leaves the recovery channel unchanged and requires no additional physical gate. This freedom does not remove relative phases between singular-value sectors. Let $r_k^{[a,b]}$ be an optimizing real polynomial, $p_k^{[a,b]}$ a corresponding complex polynomial, and $\vec\theta_k^{~[a,b]}$ a realizing angle sequence. If all nonzero scaled singular values lie in $[a,b]$, then
\begin{equation}
    \begin{aligned}
    \left|\sum_i\lambda_i^2 p_k^{[a,b]}(x_i)\right|
    &\geq \sum_i\lambda_i^2 r_k^{[a,b]}(x_i)\\
    &\geq 1-\delta^{[a,b]}(k),
    \end{aligned}
\end{equation}
using $\sum_i\lambda_i^2=1$. The corresponding infidelity $\mathcal I_k^{[a,b]}$ therefore satisfies
\begin{equation}
    \mathcal I_k^{[a,b]}
    \leq 1-\left(1-\delta^{[a,b]}(k)\right)^2
    \leq 2\delta^{[a,b]}(k).
\end{equation}

The tightest interval containing the nonzero scaled singular values is $[a,b]=[x_{\min},x_{\max}]$. Setting $b=1$ in Eq.~\eqref{eq:optimal_polynominal_minmax} requires only the minimum singular value $x_{\min}$, at the cost of a potentially larger worst-case deviation. For this case, an exact analytic phase sequence is available at $k=2$; higher-order constructions are obtained through numerical polynomial optimization and phase synthesis~\cite{Low2016Methodology}. Fig.~\ref{fig:optimal_poly} compares optimization on $[0.2,1]$ with optimization on $[0.2,0.6]$, demonstrating that the worst-case deviation increases with the larger interval and decreases with the larger iteration number $k$.

For comparison, the optimal fixed-point search~\cite{Yoder2014FixedPoint} maximizes the worst-case success probability. In our notation, this is equivalent to
\begin{equation}\label{eq:fixed_point_objective}
    \text{fixed point:}\quad \max_p \min_{x\in[a,1]} |p(x)|.
\end{equation}
Its solution is expressed through Chebyshev polynomials, with an analytic angle sequence for every $k$. This objective does not require phase alignment between different singular values: large $|p(x_i)|$ alone does not ensure a large coherent average $\left|\sum_i\lambda_i^2p(x_i)\right|$. It therefore guarantees search success but does not by itself guarantee an accurate Petz recovery.

Given a target polynomial satisfying the QSP completion conditions, a corresponding angle sequence can be computed efficiently using established classical synthesis methods~\cite{Gilyen2019Quantum,Haah2019Product,Dong2021Efficient,Martyn2021Grand}. In our numerical implementation, real target polynomials $r$ are represented in the Chebyshev basis, and symmetric QSP phases are fitted by nonlinear least squares at Chebyshev nodes, gradually increasing the target amplitude from zero to its desired value. After a common phase choice, the real part $\operatorname{Re}[e^{-i\phi}p(x)]$ is checked against $r(x)$, together with the QSP unitarity condition. This synthesis accuracy is separate from the approximation error of the chosen target polynomial.

\subsection{Convergence and Iteration Complexity}\label{sec:convergence}
At fixed iteration count $k$, define the optimal diamond distance as $\Delta^{\opt}(k)=\min_{\vec\theta}\Delta(\vec\theta)$. This minimizes the actual channel distance, whereas spectrum-weighted optimization minimizes its upper bound. Choosing the interval $[a,1]$ with $0<a\leq x_{\min}$ and combining the polynomial-error estimate above with Eq.~\eqref{eq:diamond_norm_bound} gives
\begin{equation}
    \Delta^{\opt}(k) \leq \sqrt{2\|\sigma^{-1}\|_{\infty}\delta^{[a,1]}(k)}.
\end{equation}
For $b=1$, Eq.~\eqref{eq:optimal_polynominal_minmax} is related to the classical best uniform approximation of the sign function on $[-1,-a]\cup[a,1]$~\cite{Eremenko2007Uniform,Eremenko2011Polynomials}, which gives
% .. Eremenko and Yuditskii derived its precise asymptotics for symmetric intervals and subsequently generalized the analysis to asymmetric intervals
%Let $L_k(a)$ be the classical minimax error for degree at most $2k+1$, and let $r_k$ be the corresponding odd extremal polynomial. The polynomial $r_k/(1+L_k(a))$ is bounded by one on $[-1,1]$ and has minimum $(1-L_k(a))/(1+L_k(a))$ on $[a,1]$. Conversely, rescaling the real part of any feasible polynomial with deficit $\delta$ by $2/(2-\delta)$ gives classical approximation error $\delta/(2-\delta)$. Hence $\delta^{[a,1]}(k)=2L_k(a)/(1+L_k(a))$.
% Using Theorem~1 of Ref.~\cite{Eremenko2007Uniform}, equivalently Eq.~(1.1) of Ref.~\cite{Eremenko2011Polynomials} with $A=1/a$ and $m=k$, gives, for fixed $0<a<1$,
\begin{equation}\label{eq:delta_asymptotic}
    \delta^{[a, 1]}(k) \sim \frac{2(1-a)}{\sqrt{\pi a k}}\left(\frac{1-a}{1+a}\right)^k
\end{equation}
for large $k$.

\begin{figure}
    \centering
    \includegraphics[width=\linewidth]{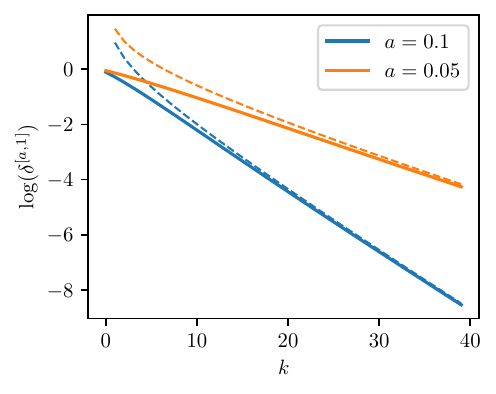}
    \caption{Logarithm of the approximation error $\delta^{[a,1]}$ as a function of the iteration number $k$ for different values of $a$. The dashed lines represent the asymptotic expressions in Eq.~\eqref{eq:delta_asymptotic}.}
    \label{fig:delta_k}
\end{figure}

Fig.~\ref{fig:delta_k} compares the numerically optimized error with this asymptotic expression for different values of $a$. For fixed $0<a<1$, the asymptotic decay rate is approximately $2a$ when $a$ is small. Taking $a=x_{\min}$, we obtain, as $k\to\infty$,
\begin{equation}
    \Delta^{\opt}(k)=\mathcal O\!\left(k^{-1/4}\left(\frac{1-x_{\min}}{1+x_{\min}}\right)^{k/2}\right)
    =\mathcal O(e^{-x_{\min}k}).
\end{equation}
Thus, for fixed $\sigma$ and $a=x_{\min}>0$, the optimal diamond distance between the composite encoding--recovery channels has an exponentially decaying upper bound. For a target diamond-distance error $\varepsilon$, this gives a sufficient iteration count scaling as $k=\mathcal O(x_{\min}^{-1}\log(1/\varepsilon))$ as $\varepsilon\to0$, with the reference state and spectrum fixed. We expect that by setting $b$ to $x_{\max}$, the decay rate will be larger and depend on both the minimum and maximum singular values, and the bound will be tighter.

\subsection{Explicit Block-Encoding Circuits}
\label{sec:block_encoding_circuits}
While the block encoding of a general operator $\rowE*{\sigma}$ is not unique, we show that the Stinespring dilation of the encoding channel $\E$ provides two natural forms. Using the dilation of the encoding channel, we can write
\begin{equation}
    \begin{aligned}
    \rowE{\sigma} &= \sum_i \bra{i}_C \otimes \left(\bra{i}_C U\ket{0}_B\sigma^{\frac{1}{2}}\right) =
    \begin{tikzpicture}[baseline={([yshift=-0.5ex]current bounding box.center)}]
    \pic[unitary/B label=$\ket{0}_B$, unitary/D label=$D$] (u1) at (0,0) {unitary};
    \draw[thick] (u1-A) -- ++(0,-0.25) node[
        rectangle,
        draw=black,
        fill=white!5,
        very thick,
        minimum size=.2cm
      ] {$\sigma^{\frac{1}{2}}$} --++(0,-0.5) coordinate (sig) node[below] {$A$};
    \draw[thick] (u1-C) -| ([xshift=-0.5cm]sig) node[below] {$C$};
\end{tikzpicture}.
    \end{aligned}
\end{equation}

The first block-encoding form reads
\begin{equation}\label{eq:block_encoding_form1}
    \text{(I):}\quad \rowE*{\sigma} = \sqrt{d_A d_C}
        \begin{tikzpicture}[baseline={([yshift=-0.5ex]current bounding box.center)}]
            \pic[unitary/B label=$\bra{0}_B$, unitary/D label=$D$] (u1) at (0,0) {unitaryd};
            \draw[thick] (u1-A) -- ++(-0.5, 0) node[midway, circle, fill, inner sep=1.5pt] {}-- ++ (0, -1.5) -- ++(-1, 0) node[midway,
                rectangle,
                draw=black,
                fill=white!5,
                very thick,
                minimum size=.2cm
            ] {$\sigma^{\frac{1}{2}}$} coordinate (tmp) -- (tmp |- u1-A) node[above] {$A$};
            \draw[thick] (u1-C) --++(0, -0.75) -- ++(-2, 0) coordinate (tmp) node[midway, circle, fill, inner sep=1.5pt] {} -- (tmp|-u1-A) node[above] {$C$};
            \draw[dashed] (-3,-0.5)--++ (4,0);
            \draw[dashed] (-3,0.5)--++ (4,0);
        \end{tikzpicture}.
\end{equation}
We can read off the initial state $\ket{s}$, the target state $\ket{t}$, and the unitary operator $W$ from the diagram above:
\begin{equation}
    \begin{aligned}
    \alpha = \sqrt{d_A d_C}, \quad W = I_C \otimes I_A \otimes I_{A'} \otimes U^\dagger,\\
    \ket{s} = \ket{\text{EPR}}_{C C'} \otimes \ket{\sigma}_{A,A'}, \quad \ket{t} = \ket{\text{EPR}}_{A' A''} \otimes \ket{0}_B.
    \end{aligned}
\end{equation}
Here $\ket{\sigma}_{AA'}$ denotes the canonical purification of $\sigma$. The total dimension of the extended Hilbert space is given by $d_{\mathrm{ext}}^{(\mathrm{I})} = d_A^2 d_C^2 d_D$.

The second block-encoding form reads
\begin{equation}\label{eq:block_encoding_form2}
    \text{(II):}\quad \rowE*{\sigma} = \sqrt{d_D} \begin{tikzpicture}[baseline={([yshift=-0.5ex]current bounding box.center)}]
        \pic[unitary/B label=$\bra{0}_B$, unitary/C label=$C$] (u1) at (0,0) {unitarystar};
        \draw[thick] (u1-A) --++ (0, -0.5) -- ++ (1, 0) coordinate (tmp) node[midway, rectangle, draw=black, fill=white!5, very thick, minimum size=.2cm] {$\sigma^{\frac{1}{2}}$} -- (tmp |- u1-C) node[above] {$A$};
        \draw[thick] (u1-D)-- ++ (-1, 0) node[midway, circle, fill, inner sep=1.5pt] {} coordinate (tmp) -- (tmp |- u1-B) node[below] {$D$};
        \draw[dashed] (-2,-0.5)--++ (4,0);
        \draw[dashed] (-2,0.5)--++ (4,0);
    \end{tikzpicture}.
\end{equation}
The initial state, target state, and unitary operator are given by
\begin{equation}
    \begin{aligned}
    \alpha =\sqrt{d_D},\quad W = I_D \otimes U^* \otimes I_A, \\
    \ket{s} = \ket{0}_B \otimes \ket{\sigma}_{A A'}, \quad \ket{t} = \ket{\text{EPR}}_{DD'}.
    \end{aligned}
\end{equation}
The total dimension of the extended Hilbert space is given by $d_{\mathrm{ext}}^{(\mathrm{II})} = d_A d_C d_D^2$.

Comparing the two forms, we see that 
\begin{equation}\label{eq:block_encoding_ratio}
    \left(\frac{\alpha^{(\text{I})}}{\alpha^{(\text{II})}}\right)^2 = \frac{\dim(\H_{\mathrm{ext}}^{(\mathrm{I})})}{\dim(\H_{\mathrm{ext}}^{(\mathrm{II})})} = \frac{d_A d_C}{d_D}.
\end{equation}
Form I has both a smaller normalization factor $\alpha$ and a smaller extended space when $d_D>d_A d_C$, whereas form II has these advantages when $d_D<d_A d_C$. A smaller $\alpha$ increases $x_{\min}=\lambda_{\min}/\alpha$ and improves the iteration bound in Sec.~\ref{sec:convergence}. This comparison concerns normalization and space; the total circuit cost also depends on preparing $\ket{\sigma}$ and implementing the required unitaries. Form I requires access to $U^\dagger$ and its inverse, while form II requires access to $U^*$ and its inverse.

For $\sigma=I_A/d_A$, type I can be reduced to type I$'$:
\begin{equation}\label{eq:block_encoding_form1_mixed}
    \text{(I$'$):}\quad \rowE*{\sigma}=\sqrt{\frac{d_C}{d_A}}\,
    \begin{tikzpicture}[baseline={([yshift=-0.5ex]current bounding box.center)}]
        \pic[unitary/A label=$A$, unitary/B label=$\bra{0}_B$, unitary/D label=$D$] (u1) at (0,0) {unitaryd};
        \draw[thick] (u1-C) -- ++(0,-0.55) -- ++(-1,0)
            node[midway, circle, fill, inner sep=1.5pt] {} coordinate (tmp)
            -- (tmp |- u1-A) node[above] {$C$};
        \draw[dashed] (-1.8,-0.5)--(1.1,-0.5);
        \draw[dashed] (-1.8,0.5)--(1.1,0.5);
    \end{tikzpicture}.
\end{equation}
The corresponding parameters are
\begin{equation}
    \begin{aligned}
    \alpha = \sqrt{\frac{d_C}{d_A}},\quad W = I_C\otimes U^\dagger,\\
    \ket{s} = \ket{\text{EPR}}_{CC'},\qquad \ket{t}=\ket{0}_B.
    \end{aligned}
\end{equation}
The extended-space dimension is $d_{\mathrm{ext}}^{(\mathrm{I}')}=d_C^2d_D$. Compared with type I, $\alpha$ and $d_{\mathrm{ext}}$ decrease by factors of $d_A$ and $d_A^2$, respectively.
Comparing type I$'$ with type II gives
\begin{equation}\label{eq:block_encoding_ratio_mixed}
    \left(\frac{\alpha^{(\mathrm{I}')}}{\alpha^{(\mathrm{II})}}\right)^2
    =\frac{d_{\mathrm{ext}}^{(\mathrm{I}')}}{d_{\mathrm{ext}}^{(\mathrm{II})}}
    =\frac{d_C}{d_A d_D}.
\end{equation}
Thus, for maximally mixed $\sigma$, type I$'$ has smaller normalization and space requirements when $d_D>d_C/d_A$, while type II has these advantages when $d_D<d_C/d_A$.

We also note that more efficient block-encoding forms may be possible for specific encoding channels. We will provide an example of this in Sec.~\ref{sec:HP} and leave further exploration in this direction to future work.

\section{Applications to Quantum Many-Body Systems}
We now test our polar-based Petz recovery algorithm on quantum many-body decoding problems. We consider two encoding channels: a Haar-random unitary encoding channel and the Hayden--Preskill encoding channel, which is closely related to the random unitary encoding channel but has additional entanglement structure. In both examples, we choose the maximally mixed reference state $\sigma=I_A/d_A$ and $n_A=1$ logical qubit. Here $n$ counts the qubits acted on by $U$, excluding the additional reference register in the Hayden--Preskill setup. We compare three optimized angle sequences from Sec.~\ref{sec:angle_finding} and the standard Grover sequence:
\begin{itemize}
    \item $\vec\theta_k^{\mathrm w}$ from Eq.~\eqref{eq:optimal_polynomial}
    \item $\vec\theta_k^{~[x_{\min},x_{\max}]}$ from Eq.~\eqref{eq:optimal_polynominal_minmax}
    \item $\vec\theta_k^{~[x_{\min},1]}$
    \item $\theta_i =\pi$ from the standard Grover search method.
\end{itemize}
The first angle sequence requires knowledge of the full spectrum of $\rowE*{\sigma}$. The second angle sequence requires knowledge of both the minimum and maximum singular values, while the third angle sequence only requires the minimum singular value. The last angle sequence is independent of the spectral information.
We will compare the entanglement fidelity of the approximate Petz recovery channels corresponding to the different angle sequences with that of the ideal Petz recovery channel as a function of the number of iterations $k$. The approximate recovery fidelities are calculated using the partial-trace norm expressions derived in Appendix~\ref{sec:approximate_petz_fidelity}.

\subsection{Recovery from Haar-Random Unitary Encoding}
In the Haar-random code~\cite{Klesse2007Approximate,Hayden2008Decoupling,Ma2025Haar,Sommers2026Spectral}, the auxiliary register $B$ is initialized in $\ket{0}_B$, and the channel isometry is
\begin{equation}\label{eq:haar_encoding_isometry}
    \colE{I}=U(I_A\otimes\ket{0}_B)=
    \begin{tikzpicture}[baseline={([yshift=-0.5ex]current bounding box.center)}]
        \pic[unitary/A label=$A$, unitary/B label=$\ket{0}_B$, unitary/C label=$C$, unitary/D label=$D$] (u1) at (0,0) {unitary};
    \end{tikzpicture},
\end{equation}
where the encoding unitary $U$ is drawn from the Haar measure. Tracing out $C$ gives the encoding channel $\E(\rho)=\tr_C[\colE{I}\rho\colE*{I}]$, whose output $D$ is available to the decoder.
Haar-random encoding serves as an idealized model of strongly scrambled quantum information~\cite{Xu2024Scrambling}, where one attempts to recover the initial information from a subsystem ($D$) of the output.
Related work connects the entanglement structure of tripartite Haar-random states to limitations on logical-operator reconstruction from individual subsystems~\cite{Li2025Tripartite}.

\begin{figure}[tbp]
    \centering
    \includegraphics{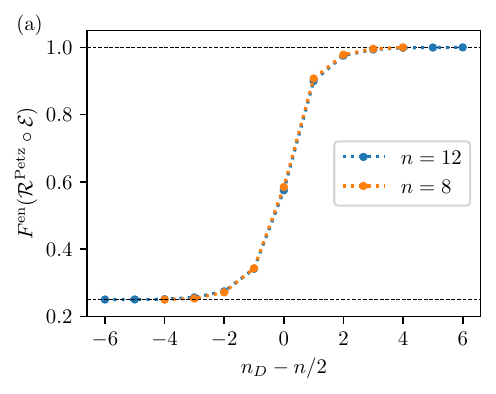}
    \includegraphics{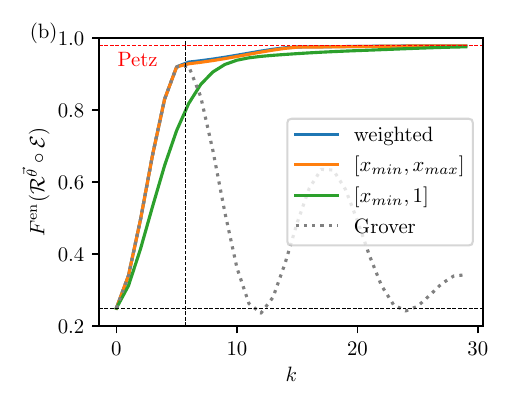}
    \caption{Petz recovery for Haar-random unitary encoding of one logical qubit ($n_A=1$). Results use a single Haar-random realization for each system size. (a) Ideal Petz entanglement fidelity versus $n_D-n/2$ where $n$ is the number of qubits acted on by $U$. The black dashed lines mark fidelities $1/4$ and $1$. (b) Entanglement fidelity of the approximate recovery versus iteration count $k$ for $n=8$, $n_D=6$, using the generic type I block encoding in Eq.~\eqref{eq:block_encoding_form1}. The vertical dashed line marks the Grover estimate $k^*$ in Eq.~\eqref{eq:grover_iteration}.}
    \label{fig:Petz_ent_fid}
\end{figure}

To provide a baseline for our algorithm, we first study the ideal Petz entanglement fidelity $\Fent(\petzChannel{\sigma} \circ \E)$ as a function of $n_D$, the number of qubits in the output subsystem \(D\), as shown in Fig.~\ref{fig:Petz_ent_fid}. The fidelity is computed using the ideal Petz recovery map, which is independent of the iteration number \(k\). Consistent with decoupling-based recovery guarantees~\cite{Hayden2008Decoupling}, the fidelity increases with \(n_D\) and rapidly approaches unity when \(n_D\) exceeds the combined size of the information subsystem \(A\) and the environment \(C\). The displayed system sizes, $n=8$ and $n=12$, give similar fidelity curves when plotted against $n_D-n/2$. This suggests a sharpening transition near $n_D/n=1/2$ as the system size increases.

Now we study the approximate recovery channel produced by our algorithm using the angle sequences described above and compare its entanglement fidelity with that of the ideal Petz recovery map as a function of the iteration number \(k\). As discussed in Sec.~\ref{sec:angle_finding}, the expected $k$ depends on the spectrum $\lambda$ of $\rowE{\sigma}$ and the block-encoding normalization $\alpha$. Since $\sum_i \lambda_i^2=1$, we have
\begin{equation}
    \overline{\lambda^2} = \frac{1}{\min(d_A d_C, d_D)},
\end{equation}
where the overline denotes the uniform average over the nonzero singular values, and $\min(d_A d_C, d_D)$ is the smaller dimension of $\rowE{\sigma}$. For the following analysis, we compare the two generic block encodings in Eqs.~\eqref{eq:block_encoding_form1} and \eqref{eq:block_encoding_form2}. Among these, type I is preferred when $d_A d_C < d_D$, and type II when $d_A d_C > d_D$, giving $\alpha^2=\min(d_A d_C,d_D)$. Therefore, the mean squared scaled singular value is
\begin{equation}\label{eq:haar_x}
    \overline{x^2} =\frac{1}{\min^2(d_A d_C, d_D)},
\end{equation}
which is exponentially small unless $n_D \sim 1 $ or $n_D \sim n$.
Using Weingarten calculus for Haar-random unitaries~\cite{Collins2006Integration}, we also estimate the standard deviation $\Std(x^2)$ in Appendix~\ref{sec:sv}. The spectral width is comparable to the mean near $d_A d_C=d_D$, while the relative width $\Std(x^2)/\overline{x^2}$ decreases exponentially as $n_D$ increases beyond this crossover.

We choose $n=8$ and $n_D = 6$ for this comparison, and since $d_A d_C < d_D$, we use the generic type I block encoding in Eq.~\eqref{eq:block_encoding_form1}. 
The results are shown in Fig.~\ref{fig:Petz_ent_fid}(b) for the four angle sequences described above.
The fidelity from the standard Grover search method increases rapidly at first and reaches a maximum around the vertical line marking the estimated first peak for Grover search, calculated using the mean value of the spectrum:
\begin{equation}\label{eq:grover_iteration}
    k^* = \frac{\pi}{4\,\arcsin\!\left(\sqrt{\overline{x^2}}\right)}-\frac{1}{2}.
\end{equation}
Near the estimated first peak $k^*$, the entanglement fidelity reached by the standard Grover search method is still below that of the ideal Petz recovery map, due to the finite width of the spectrum. Beyond this first peak, the fidelity starts to decrease and oscillate due to the overshooting effect inherent in the Grover search algorithm. The spectrum-weighted and interval-optimized sequences avoid this overshooting in the displayed results and gradually approach the ideal Petz recovery fidelity. The sequences $\vec\theta_k^{\mathrm w}$ and $\vec\theta_k^{~[x_{\min},x_{\max}]}$ perform similarly and overlap with Grover search for $k<k^*$. In this example, optimization over the tight spectral interval achieves performance comparable to spectrum-weighted optimization. The angle sequence $\vec\theta_k^{~[x_{\min},1]}$ also eventually converges to the ideal Petz recovery fidelity, but more slowly, because it additionally constrains $r(x)$ to be close to $1$ for $x>x_{\max}$. These results illustrate convergence toward the ideal Petz entanglement fidelity, consistent with the channel-error bound derived above.

For this Haar-random unitary encoding problem, the small scaled singular values in Eq.~\eqref{eq:haar_x} typically require an exponentially large number of iterations for these generic block encodings to approximate the Petz recovery map, except when $n_D$ is very small or close to $n$. Additional structure in the encoding channel can nevertheless yield larger singular values or a block encoding with a substantially smaller $\alpha$, reducing the iteration count. The Hayden--Preskill protocol, discussed next, provides such an example through entanglement assistance.

\subsection{Recovery in the Hayden--Preskill Protocol}\label{sec:HP}
\subsubsection{Encoding and Ideal Petz Recovery}
In this section, we apply our algorithm to the Hayden--Preskill (HP) encoding channel~\cite{Hayden2007Black} and compare it with the ideal Petz recovery map as well as other decoding protocols. The HP protocol is a paradigmatic example of information recovery in quantum many-body systems, where the goal is to retrieve information that has been scrambled by a chaotic unitary evolution. In this setup, $\ket{0}_B$ is replaced by a maximally entangled state \(\ket{\text{EPR}}_{BB'}\) between the auxiliary subsystems $B$ and $B'$. The unitary $U$ acts only on the logical space $A$ and the subsystem $B$, while $B'$ passes directly through to the output. The channel isometry $\colE{I}$ reads
\begin{equation}\label{eq:hp_encoding_isometry}
    \begin{aligned}
    \colE{I}&=(U\otimes I_{B'})(I_A\otimes\ket{\text{EPR}}_{BB'})\\
    &=\begin{tikzpicture}[baseline={([yshift=-0.5ex]current bounding box.center)}]
        \pic[unitary/A label=$A$, unitary/C label=$C$, unitary/D label=$D$] (u1) at (0,0) {unitary};
        \draw[thick] (u1-B) -| ([xshift=1cm]u1-D) node[above]{$B'$};
        \filldraw[black] ([xshift=0.5cm]u1-B) circle (2pt);
    \end{tikzpicture}.
    \end{aligned}
\end{equation}
In the above diagram, both subsystems $D$ and $B'$ are part of the encoding subspace where the decoding is applied. We write $d_{B'}=d_B$ and $n_{B'}=n_B$; $n=n_A+n_B=n_C+n_D$ counts the qubits acted on by $U$, excluding the additional register $B'$.
% One can view the HP encoding channel as a special case of the general construction by identifying $U$ as $U\otimes I_{B'}$, $\ket{0}_B$ as $\ket{\text{EPR}}_{BB'}$, and the general output space $\H_D$ as $\H_D\otimes\H_{B'}$.

\begin{figure}[tbp]
    \centering
    \includegraphics{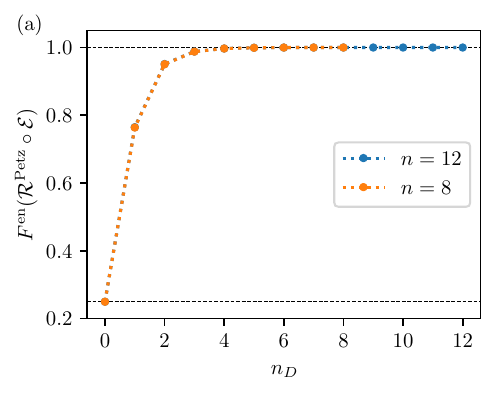}
    \includegraphics{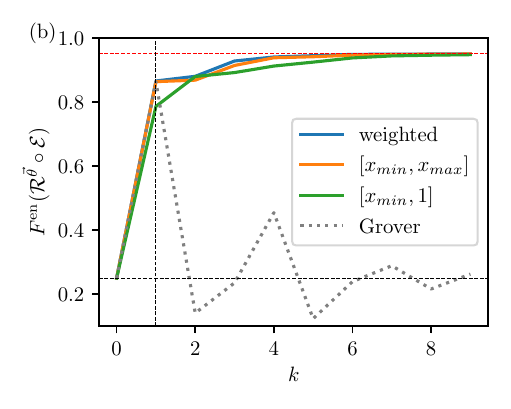}
    \caption{Petz recovery in the Hayden--Preskill protocol with one logical qubit ($n_A=1$) and a Haar-random encoding unitary. Results use a single Haar-random realization for each system size. (a) Ideal Petz entanglement fidelity versus the number of output qubits $n_D$. Here $n$ counts the qubits acted on by $U$, excluding the entangled register $B'$, which is also available to the decoder. Near-unit fidelity is obtained with only a few qubits in $D$. (b) Entanglement fidelity versus iteration count $k$ for $n=8$, $n_D=2$, using the reduced (II$'$)-HP block encoding in Eq.~\eqref{eq:block_encoding_form2_reduced_hp}. One Grover iteration gives a large initial improvement, while the optimized sequences continue toward the Petz benchmark.}
    \label{fig:HP_Petz_ent_fid}
\end{figure}

We again draw $U$ from the Haar ensemble, but now with the additional entanglement structure of the HP protocol. We first compute the entanglement fidelity of the ideal Petz recovery map as a function of the number of qubits in the output subsystem \(D\), as shown in Fig.~\ref{fig:HP_Petz_ent_fid}. As a hallmark of the HP protocol, the fidelity approaches unity from the baseline $1/4$ with only a few qubits in $D$. Here the decoder accesses both $D$ and the entangled register $B'$, so near-unit recovery from a small $D$ benefits from side information absent in the unassisted Haar-random encoding. The displayed system sizes, $n=8$ and $n=12$, give similar fidelity curves, as in the random encoding example.

\subsubsection{Reduced Block Encoding and Recovery Performance}
Now we study the singular-value distribution and block encoding of $\rowE*{\sigma}$ for the HP encoding channel. The average singular value squared is
\begin{equation}
    \overline{\lambda^2} = \frac{d_D}{d_B}\frac{1}{\min(d_A^2, d_D^2)}.
\end{equation}
The singular values are still exponentially small unless $n_D\sim n_B$. However, the specific structure of the HP encoding channel allows for a much more efficient block encoding of $\rowE*{\sigma}$ than is possible for a general encoding channel, as we show next.

% \begin{equation}
%      \rowE*{\sigma} = \begin{tikzpicture}[baseline={([yshift=-0.5ex]current bounding box.center)}]
%         \pic (u1) at (0,0) {unitaryd};
%         \draw[thick] (u1-B) -- ++ (0.5, 0) node[midway, circle, fill, inner sep=1.5pt] {} coordinate (tmp) -- (tmp |- u1-D) node[below] {$B$};
%         \draw[thick] (u1-A) -- ++ (0, 0.25)  node[ rectangle, draw=black, fill=white!5, very thick, minimum size=.2cm] {$\sigma^{\frac{1}{2}}$} --++ (0, 0.5) coordinate (A) node[above] {$A$};
%         \draw[thick] (u1-C) -| ([xshift=-0.5cm]A);
%     \end{tikzpicture}
% \end{equation}

We start by considering the two generic block-encoding forms of $\rowE*{\sigma}$ introduced in Eqs.~\eqref{eq:block_encoding_form1} and \eqref{eq:block_encoding_form2}. The type I form is
\begin{equation}
    \text{(I)-HP:}\quad \rowE*{\sigma}=
     \sqrt{d_A d_C} 
        \begin{tikzpicture}[baseline={([yshift=-0.5ex]current bounding box.center)}]
            \pic[unitary/D label=$D$] (u1) at (0,0) {unitaryd};
            \draw[thick] (u1-A) -- ++(-0.5, 0) node[midway, circle, fill, inner sep=1.5pt] {}-- ++ (0, -1.5) -- ++(-1, 0) node[midway,
                rectangle,
                draw=black,
                fill=white!5,
                very thick,
                minimum size=.2cm
            ] {$\sigma^{\frac{1}{2}}$} coordinate (tmp) -- (tmp |- u1-A) node[above] {$A$};
            \draw[thick] (u1-C) --++(0, -0.75) -- ++(-2, 0) coordinate (tmp) node[midway, circle, fill, inner sep=1.5pt] {} -- (tmp|-u1-A) node[above] {$C$};
            \draw[thick] (u1-B) -- ++ (0.5, 0) node[midway, circle, fill, inner sep=1.5pt] {} coordinate (tmp) -- (tmp |- u1-D) node[below] {$B'$};
            \draw[dashed] (-3,-0.5)--++ (4.5,0);
            \draw[dashed] (-3,0.5)--++ (4.5,0);
        \end{tikzpicture}.
\end{equation}
For $\sigma=I_A/d_A$, the type I$'$ simplification gives
\begin{equation}\label{eq:block_encoding_form1_mixed_hp}
    \text{(I$'$)-HP:}\quad \rowE*{\sigma}=\sqrt{\frac{d_C}{d_A}}\,
    \begin{tikzpicture}[baseline={([yshift=-0.5ex]current bounding box.center)}]
        \pic[unitary/A label=$A$, unitary/D label=$D$] (u1) at (0,0) {unitaryd};
        \draw[thick] (u1-C) -- ++(0,-0.55) -- ++(-1,0)
            node[midway, circle, fill, inner sep=1.5pt] {} coordinate (tmp)
            -- (tmp |- u1-A) node[above] {$C$};
        \draw[thick] (u1-B) -- ++(0.5,0)
            node[midway, circle, fill, inner sep=1.5pt] {} coordinate (tmp)
            -- (tmp |- u1-D) node[below] {$B'$};
        \draw[dashed] (-1.8,-0.5)--(1.3,-0.5);
        \draw[dashed] (-1.8,0.5)--(1.3,0.5);
    \end{tikzpicture}.
\end{equation}
The type II form is
\begin{equation}
    \text{(II)-HP:}\quad \rowE*{\sigma} = \sqrt{d_D d_B} \begin{tikzpicture}[baseline={([yshift=-0.5ex]current bounding box.center)}]
        \pic[unitary/C label=$C$] (u1) at (0,0) {unitarystar};
        \draw[thick] (u1-A) --++ (0, -0.5) -- ++ (1, 0) coordinate (tmp) node[midway, rectangle, draw=black, fill=white!5, very thick, minimum size=.2cm] {$\sigma^{\frac{1}{2}}$} -- (tmp |- u1-C) node[above] {$A$};
        \draw[thick] (u1-D) --++(0,0.3)-- ++ (-1.5, 0) node[midway, circle, fill, inner sep=1.5pt] {} coordinate (tmp) -- (tmp |- u1-B) node[below] {$D$};
        \draw[thick] (u1-B) --++ (-0.5, 0) node[midway, circle, fill, inner sep=1.5pt] {} coordinate (tmp) -- (tmp |- u1-D) -- ++ (-0.5,0) node[midway, circle, fill, inner sep=1.5pt] {} coordinate (tmp) -- (tmp |- u1-B) node[below] {$B'$};
        \draw[dashed] (-2.2,-0.5)--++ (4,0);
        \draw[dashed] (-2.2,0.5)--++ (4,0);
    \end{tikzpicture}.
\end{equation}
The normalization factors in both forms are consistent with the random unitary encoding protocol and are generally exponentially large. Therefore, the scaled singular value $x$ is still exponentially small.

However, type II can be further reduced to a more efficient type II$'$ form:
\begin{equation}\label{eq:block_encoding_form2_reduced_hp}
    \text{(II$'$)-HP:}\quad \rowE*{\sigma} = \sqrt{d_D/d_B} \begin{tikzpicture}[baseline={([yshift=-0.5ex]current bounding box.center)}]
        \pic[unitary/B label =$B'$, unitary/C label=$C$] (u1) at (0,0) {unitarystar};
        \draw[thick] (u1-A) --++ (0, -0.5) -- ++ (1, 0) coordinate (tmp) node[midway, rectangle, draw=black, fill=white!5, very thick, minimum size=.2cm] {$\sigma^{\frac{1}{2}}$} -- (tmp |- u1-C) node[above] {$A$};
        \draw[thick] (u1-D) -- ++ (-0.5, 0) node[midway, circle, fill, inner sep=1.5pt] {} coordinate (tmp) -- (tmp |- u1-B) node[below] {$D$};
        \draw[dashed] (-1.2,-0.5)--++ (3,0);
        \draw[dashed] (-1.2,0.5)--++ (3,0);
    \end{tikzpicture}.
\end{equation}
The normalization factor is now $\alpha^{(\mathrm{II}')\text{-HP}}=\sqrt{d_D/d_B}$, smaller than that of the generic type II form by a factor of $d_B$ and no larger than that of the generic type I form, with equality when $d_C=1$. For maximally mixed $\sigma$, comparison with type I$'$ applied to the HP channel gives $\bigl(\alpha^{(\mathrm{II}')\text{-HP}}\bigr)^2/\bigl(\alpha^{(\mathrm{I}')\text{-HP}}\bigr)^2=d_A^2/d_C^2$, thus reduced type II has smaller normalization when $d_C>d_A$, ties when $d_C=d_A$, and has larger normalization when $d_C<d_A$. For the reduced type II construction, the average scaled singular value squared is
\begin{equation}
    \overline{x^2} = \frac{1}{\min(d_A^2, d_D^2)}.
\end{equation}
As shown in Appendix~\ref{sec:sv}, the standard deviation $\Std(x^2)$ is comparable to the average $\overline{x^2}$ when $d_D=d_A$ but decreases exponentially as $n_D$ increases.
Since the number of logical qubits $n_A$ is typically small ($n_A =1$ for the current example), the scaled singular value is no longer exponentially small and the number of iterations needed to achieve Petz fidelity is significantly reduced. 
% This clearly demonstrates the advantage of using entangled states in quantum error correction, the idea behind entanglement-assisted quantum error correction.

Now we compare the entanglement fidelity obtained using our algorithm and the reduced type II block encoding with that of the ideal Petz recovery map. The results are shown in Fig.~\ref{fig:HP_Petz_ent_fid}(b) for the four different angle sequences. As expected, the number of iterations required to achieve Petz fidelity is significantly reduced compared with that for the random unitary encoding protocol.
For $n_A=1$ and $n_D\geq1$, $\overline{x^2}=1/4$. When the spectrum is concentrated near $x=1/2$, Eq.~\eqref{eq:grover_iteration} predicts $k^*=1$: a single Grover iteration maps the amplitude $x=1/2$ to $1$. The finite spectral width limits the recovery fidelity, which remains below that of ideal Petz recovery in the displayed example. As in the Haar-random code, the spectrum-weighted and interval-optimized sequences improve the fidelity toward the Petz benchmark as the iteration count $k$ increases.

\subsubsection{Relation to Existing Decoders}\label{sec:existing_decoders}
The known HP decoders can be organized by the choice of block encoding, iteration count, and postselection in Algorithm~\ref{alg:petz}, providing a unified framework for understanding and improving these decoders.

At $k=0$, $\recoveryW = W$ and the corresponding polynomials are $p(x)=x$ and $q(x)=1$. The (I$'$)-HP block encoding in Eq.~\eqref{eq:block_encoding_form1_mixed_hp}, with postselection on $\ket{t}$, realizes the Petz-lite decoder~\cite{Nakayama2023Petz}:
\begin{equation}\label{eq:hp_petz_lite_decoder}
    \begin{tikzpicture}[baseline={([yshift=-0.5ex]current bounding box.center)}]
        \pic[unitary/A label=$\ket{\psi}_A$, unitary/C label=$C$] (enc) at (0,0) {unitary};
        \pic[unitary/A label=$A'$] (rec) at (0,2.7) {unitaryd};
        \draw[thick] (enc-D) -- (rec-D) node[midway, right] {$D$};
        \draw[thick] (enc-B) -- (0.5,-1.1) -- (1.7,-1.1)
            node[midway, circle, fill, inner sep=1.5pt] {}
            -- (1.7,3.9) -- (0.5,3.9)
            node[midway, circle, fill, inner sep=1.5pt] {} -- (rec-B);
        \node[right] at (1.7,1.2) {$B'$};
        \node[above] at (1.1,3.9) {$\bra{\mathrm{EPR}}_{BB'}$};
        \draw[thick] (rec-C) -- (-0.5,1.65) -- (-1.3,1.65)
            node[midway, circle, fill, inner sep=1.5pt] {}
            -- (-1.3,3.3) node[above] {$C'$};
    \end{tikzpicture}.
\end{equation}
Here the $B$ output is postselected with $B'$. Discarding the encoding environment $C$ and recovery environment $C'$ leaves the recovered state on $A'$. In the ideal limit of perfect conditional recovery, the retained $CC'$ state is an EPR pair: replacing the auxiliary EPR preparation by an input state on $C''$ then transfers that state to $C$, giving a bidirectional-teleportation interpretation related to Ref.~\cite{Vikram2026Bidirectional}. 

On the other hand, the type (II$'$)-HP circuit in Eq.~\eqref{eq:block_encoding_form2_reduced_hp}, with the same $k=0$ postselection, gives the stochastic Yoshida--Kitaev decoder~\cite{Yoshida2017Efficient}:
\begin{equation}\label{eq:hp_stochastic_yk_decoder}
    \begin{tikzpicture}[baseline={([yshift=-0.5ex]current bounding box.center)}]
        \pic[unitary/A label=$\ket{\psi}_A$, unitary/C label=$C$] (enc) at (-1.2,0) {unitary};
        \pic[unitary/C label=$C'$] (rec) at (1.2,0) {unitarystar};
        \draw[thick] (enc-B) -- (-0.7,-1.1) -- (0.7,-1.1)
            node[midway, circle, fill, inner sep=1.5pt] {} -- (rec-B);
        \draw[thick] (rec-A) -- (1.7,-1.1) -- (2.6,-1.1)
            node[midway, circle, fill, inner sep=1.5pt] {}
            -- (2.6,0.6) node[above] {$A'$};
        \draw[thick] (enc-D) -- (-0.7,1.2) -- (0.7,1.2)
            node[midway, circle, fill, inner sep=1.5pt] {} -- (rec-D);
        \node[above] at (0,1.2) {$\bra{\mathrm{EPR}}_{DD'}$};
    \end{tikzpicture}.
\end{equation}
Here $DD'$ is postselected onto an EPR state. Discarding $CC'$ leaves the recovered state on $A'$. The postselected constructions in Eqs.~\eqref{eq:hp_petz_lite_decoder} and \eqref{eq:hp_stochastic_yk_decoder} differ only by an overall scalar factor. They therefore produce identical normalized recovered states, although their success probabilities differ, consistent with Ref.~\cite{Nakayama2023Petz}. A related rearrangement of the postselected decoding circuit gives the decoding-before-encoding protocol of Ref.~\cite{Huang2026Experimental}, in which the recovered state is conditioned on a later input and Bell-measurement outcome.

Using the (II$'$)-HP construction with the Grover angle sequence $\theta_i=\pi$ yields the deterministic Yoshida--Kitaev decoder~\cite{Yoshida2017Efficient}:
\begin{equation}\label{eq:hp_deterministic_yk_decoder}
    \begin{tikzpicture}[baseline={([yshift=-0.5ex]current bounding box.center)}]
        \pic[unitary/A label=$\ket{\psi}_A$, unitary/C label=$C$] (enc) at (-1.2,0) {unitary};
        \pic (rec) at (1.2,0) {unitarystar};
        \draw[thick] (enc-B) -- (-0.7,-1.1) -- (0.7,-1.1)
            node[midway, circle, fill, inner sep=1.5pt] {} -- (rec-B);
        \draw[thick] (rec-A) -- (1.7,-1.1) -- (2.6,-1.1)
            node[midway, circle, fill, inner sep=1.5pt] {} coordinate (aout);
        \node[draw, very thick, fill=white!5, minimum width=2.1cm,
            minimum height=0.45cm] (rt) at (0,1) {$e^{i\pi\ket{t}\bra{t}}$};
        \pic (inv) at (1.2,2) {unitaryt};
        \node[draw, very thick, fill=white!5, minimum width=1.65cm,
            minimum height=0.45cm] (rs) at (2.15,3.05) {$e^{i\pi\ket{s}\bra{s}}$};
        \pic (last) at (1.2,4.1) {unitarystar};
        \draw[thick] (enc-D) -- (enc-D |- rt.south);
        \draw[thick] (enc-D |- rt.north) -- (-0.7,5) node[above] {$D$};
        \draw[thick] (rec-D) -- (rec-D |- rt.south);
        \draw[thick] (rec-D |- rt.north) -- (inv-D);
        \draw[thick] (rec-C) -- (inv-C);
        \draw[thick] (inv-B) -- (last-B);
        \draw[thick] (inv-A) -- (inv-A |- rs.south);
        \draw[thick] (inv-A |- rs.north) -- (last-A);
        \draw[thick] (aout) -- (aout |- rs.south);
        \draw[thick] (aout |- rs.north) -- (2.6,5) node[above] {$A'$};
        \draw[thick] (last-D) -- (0.7,5) node[above] {$D'$};
        \draw[thick] (last-C) -- (1.7,5) node[above] {$C'$};
        \node[draw, dashed, fit=(rt)(inv-body)(rs)(last-body),
            inner xsep=0.15cm, inner ysep=0.15cm,
            label={[font=\small]left:$k$ iterations}] {};
    \end{tikzpicture}.
\end{equation}
Here $\ket{t}=\ket{\mathrm{EPR}}_{DD'}$ and $\ket{s}=\ket{\mathrm{EPR}}_{AA'}$ are normalized EPR states. The initial $U^*$ prepares the search state, and each repeated block implements $Z_s(\pi)Z_t(\pi)$. The recovered state appears on $A'$, and the registers $DD'CC'$ are discarded without postselection. The nearly flat spectrum in the decoupling regime allows the Grover rotations to amplify the singular-vector sectors together. For our one-qubit example, the spectrum concentrated near $x=1/2$ gives the single-iteration estimate discussed above; finite spectral width limits its accuracy.

For $n_A=1$ and $k=1$, the circuit simplifies further as $\exp(i\pi\ket{s}\bra{s})$ can be related to a Pauli-$Y$ correction and a SWAP operation.
\begin{equation}
    \begin{aligned}
    e^{i\pi\ket{s}\bra{s}}&=(Y_{A''}\otimes Y_{A'})\,\mathrm{SWAP}_{A''A'}
    =\begin{tikzpicture}[baseline={([yshift=-0.5ex]current bounding box.center)}]
        \node[draw, very thick, fill=white!5, minimum size=0.4cm,
            inner sep=2pt] (yl) at (-0.45,0.95) {$Y$};
        \node[draw, very thick, fill=white!5, minimum size=0.4cm,
            inner sep=2pt] (yr) at (0.45,0.95) {$Y$};
        \draw[thick] (-0.45,0) -- (-0.45,0.15) -- (0.45,0.6) -- (yr.south);
        \draw[thick, preaction={draw=white, line width=3pt}]
            (0.45,0) -- (0.45,0.15) -- (-0.45,0.6) -- (yl.south);
        \draw[thick] (yl.north) -- (-0.45,1.45) node[above] {$A''$};
        \draw[thick] (yr.north) -- (0.45,1.45) node[above] {$A'$};
    \end{tikzpicture}.
    \end{aligned}
\end{equation}
At $k=1$, the circuit diagram becomes:
\begin{equation}\label{eq:hp_single_iteration_swap}
    \begin{tikzpicture}[baseline={([yshift=-0.5ex]current bounding box.center)}]
        \pic[unitary/A label=$\ket{\psi}_A$, unitary/C label=$C$] (enc) at (-1.2,0) {unitary};
        \pic (rec) at (1.2,0) {unitarystar};
        \draw[thick] (enc-B) -- (-0.7,-1.1) -- (0.7,-1.1)
            node[midway, circle, fill, inner sep=1.5pt] {} -- (rec-B);
        \draw[thick] (rec-A) -- (1.7,-1.1) -- (2.6,-1.1)
            node[midway, circle, fill, inner sep=1.5pt] {} coordinate (aout);
        \node[draw, very thick, fill=white!5, minimum width=2.1cm,
            minimum height=0.45cm] (rt) at (0,1) {$e^{i\pi\ket{t}\bra{t}}$};
        \pic (inv) at (1.2,2) {unitaryt};
        \node[draw, dash pattern=on 3pt off 1pt, very thick, fill=white!5, minimum size=0.4cm,
            inner sep=2pt] (yl) at (1.7,3.5) {$Y$};
        \node[draw, very thick, fill=white!5, minimum size=0.4cm,
            inner sep=2pt] (yr) at (2.6,3.5) {$Y$};
        \pic[every node/.append style={dash pattern=on 3pt off 1pt}] (last) at (1.2,4.5) {unitarystar};
        \draw[thick] (enc-D) -- (enc-D |- rt.south);
        \draw[thick] (enc-D |- rt.north) -- (-0.7,5.4) node[above] {$D$};
        \draw[thick] (rec-D) -- (rec-D |- rt.south);
        \draw[thick] (rec-D |- rt.north) -- (inv-D);
        \draw[thick] (rec-C) -- (inv-C);
        \draw[thick] (inv-B) -- (last-B);
        \draw[thick] (inv-A) -- (1.7,2.8) -- (2.6,3.2) -- (yr.south);
        \draw[thick, preaction={draw=white, line width=3pt}]
            (aout) -- (2.6,2.8) -- (1.7,3.2) -- (yl.south);
        \draw[thick] (yl.north) -- (last-A);
        \draw[thick] (yr.north) -- (2.6,5.4) node[above] {$A'$};
        \draw[thick] (last-D) -- (0.7,5.4) node[above] {$D'$};
        \draw[thick] (last-C) -- (1.7,5.4) node[above] {$C'$};
    \end{tikzpicture}.
\end{equation}
The dashed gates act only on discarded registers and can be omitted. Removing the SWAP and reading out $A''$ instead of $A'$ gives the equivalent circuit
\begin{equation}\label{eq:hp_single_qubit_teleportation}
    \begin{tikzpicture}[baseline={([yshift=-0.5ex]current bounding box.center)}]
        \pic[unitary/A label=$\ket{\psi}_A$, unitary/C label=$C$] (enc) at (-1.2,0) {unitary};
        \pic (rec) at (1.2,0) {unitarystar};
        \draw[thick] (enc-B) -- (-0.7,-1.1) -- (0.7,-1.1)
            node[midway, circle, fill, inner sep=1.5pt] {} -- (rec-B);
        \draw[thick] (rec-A) -- (1.7,-1.1) -- (2.6,-1.1)
            node[midway, circle, fill, inner sep=1.5pt] {}
            -- (2.6,3.9) node[above] {$A'$};
        \node[draw, very thick, fill=white!5, minimum width=2.1cm,
            minimum height=0.45cm] (rt) at (0,1) {$e^{i\pi\ket{t}\bra{t}}$};
        \pic (inv) at (1.2,2) {unitaryt};
        \node[draw, very thick, fill=white!5, minimum size=0.4cm,
            inner sep=2pt] (ycorr) at (1.7,3.05) {$Y$};
        \draw[thick] (enc-D) -- (enc-D |- rt.south);
        \draw[thick] (enc-D |- rt.north) -- (-0.7,3.9) node[above] {$D$};
        \draw[thick] (rec-D) -- (rec-D |- rt.south);
        \draw[thick] (rec-D |- rt.north) -- (inv-D);
        \draw[thick] (rec-C) -- (inv-C);
        \draw[thick] (inv-B) -- (0.7,3.9) node[above] {$B''$};
        \draw[thick] (inv-A) -- (ycorr.south);
        \draw[thick] (ycorr.north) -- (1.7,3.9) node[above] {$A''$};
    \end{tikzpicture}.
\end{equation}

Only the corrected output $A''$ is retained. The simplified circuit shares the structure of many-body teleportation protocols~\cite{Schuster2022ManyBody,Brown2023Quantum,Nezami2023Quantum}, with the joint EPR reflection $e^{i\pi\ket{t}\bra{t}}$ on $DD'$ replaced there by an intersystem coupling $e^{igV}$.

Our optimized sequences extend these constructions by approximating $p(x)=\sgn(x)$, up to a common phase, across the nonzero spectrum. They therefore approach the full Petz recovery, including its operator-valued inverse-square-root normalization, even when the spectrum has finite width. This connects the postselected adjoint-map decoders and Grover-based deterministic decoding to the same polar-based recovery framework.

\section{Conclusion and Outlook}
We have formulated approximate Petz recovery for a general full-rank reference state $\sigma$ using the polar representation~\cite{Ma2025Haar,Afham2026Projections,Elimelech2026Universal}
\begin{equation}
\colR{\sigma}=\sgn(\rowE*{\sigma}).
\end{equation}
This identity incorporates the inverse-square-root normalization into a single singular-value transformation, giving a direct QSVT construction for approximate quantum error correction~\cite{Gilyen2019Quantum}. The construction can be interpreted as simultaneous quantum searches with aligned phases across singular-vector sectors, connecting it to phase-preserving QSVT decoders~\cite{Utsumi2026Explicit}. By retaining the full unitary evolution and discarding the auxiliary registers, we obtain an approximate recovery channel $\R^{\vec\theta}$ without postselection.

We bound the diamond distance between the ideal Petz encoding--recovery channel $\petzChannel{\sigma}\circ\E$ and its approximation $\R^{\vec\theta}\circ\E$. The QSVT dependence enters through the spectrum-weighted coherent overlap
\begin{equation}
   f=\left|\sum_i\lambda_i^2p^{\vec{\theta}}(\lambda_i/\alpha)\right|,
\end{equation}
where $\lambda_i$ are the singular values of $\rowE*{\sigma}$, $\alpha$ is the block-encoding normalization, and $p^{\vec\theta}$ is the polynomial determined by the angle sequence. The associated loss $\mathcal I^{\vec\theta}=1-f^2$ accounts for relative phases between sectors and provides an objective for optimizing the angles to tighten the diamond-distance bound. An interval-based alternative controls the worst-case polynomial error over an interval containing the nonzero scaled singular values. Combining this approach with results on best uniform polynomial approximation~\cite{Eremenko2007Uniform,Eremenko2011Polynomials} yields an exponentially decaying upper bound. For a fixed reference state and fixed nonzero singular-value spectrum,
\begin{equation}
    \Delta^{\opt}(k)=\mathcal O(e^{-\lambda_{\min}k/\alpha}),
    \qquad k\to\infty,
\end{equation}
where $\lambda_{\min}$ is the smallest nonzero singular value. This bound quantifies the tradeoff between recovery accuracy and the number of QSVT iterations.

We apply the construction to Haar-random encoding~\cite{Klesse2007Approximate,Ma2025Haar} and the Hayden--Preskill protocol~\cite{Hayden2007Black}. Our numerical results show that the recovery fidelity approaches the ideal Petz fidelity as $k$ increases. In the entanglement-assisted Hayden--Preskill setting, the channel admits a reduced block encoding with a smaller normalization $\alpha$, lowering the iteration count needed to approach Petz recovery. The formulation also places the Yoshida--Kitaev and Petz-lite decoders~\cite{Yoshida2017Efficient,Nakayama2023Petz} within a common framework, distinguished by their block encodings, iteration counts, angle sequences, and use of postselection.

This work also opens up numerous avenues for future research. It would be interesting to compare the Petz recovery map with decoders of other quantum error correction protocols in terms of efficiency and accuracy. It would be valuable to assess the Petz recovery map for practical quantum error correction codes with realistic noise models. Moreover, it would be interesting to characterize quantum many-body systems based on whether the Petz recovery map becomes simple or exhibits a structure that can be exploited for efficient decoding. In this work, we mainly consider a random unitary $U$ drawn from the Haar measure as the encoding operation. One can instead consider structured unitaries such as $\exp(-i H t)$ plus noise, and ask which Hamiltonians and time regimes lead to a simple or efficiently implementable Petz recovery map. This could provide insights into the search for new quantum error correction codes. Lastly, the polar-based construction developed here suggests routes to more effective decoders beyond the Petz map itself.

\section{Acknowledgements}
This research was supported by the National Science Foundation (NSF) under Grant No. DMR-2443462.
Codex~(GPT-6-Astra) was used to assist with programming for polynomial optimization and polishing the manuscript.

\bibliography{references}
\appendix

\section{Completeness of the Petz Map}
\label{sec:completeness}
The Kraus operators of the composite channel $\petzChannel{\sigma} \circ \E$ are given by
\begin{equation}
    \gamma_{ij} = \sigma^{\frac{1}{2}} E_i^\dagger \E(\sigma)^{-\frac{1}{2}}E_j,
\end{equation}
where $\{E_i\}$ are the Kraus operators of the encoding channel $\E$. Now we check the completeness relation:
\begin{equation}
    \begin{aligned}
    \sum_{ij} \gamma_{ij}^\dagger \gamma_{ij} &= \sum_{ij} E_j^\dagger \E(\sigma)^{-\frac{1}{2}} E_i \sigma E_i^\dagger \E(\sigma)^{-\frac{1}{2}} E_j \\
    &=\sum_i E_i^\dagger \E(\sigma)^{-\frac{1}{2}} \E(\sigma)^{\frac{1}{2}} E_i.
    \end{aligned}
\end{equation}
Recall that $\rowE{\sigma} =\sum_i \bra{i}_C \otimes E_i \sigma^{\frac{1}{2}}$ and $\E(\sigma) = \rowE{\sigma} \rowE*{\sigma}$. Even though $\E(\sigma)^{-\frac{1}{2}} \E(\sigma)^{\frac{1}{2}}$ is not strictly equal to the identity when $\E(\sigma)$ is not full rank, from the SVD of $\rowE{\sigma}$, we have
\begin{equation}\label{eq:Esigma_identity}
\rowE*{\sigma} \E(\sigma)^{-\frac{1}{2}} \E(\sigma)^{\frac{1}{2}} = \rowE*{\sigma}.
\end{equation}
Assuming that $\sigma$ is full rank, we can rewrite the completeness relation as
\begin{equation}
    \sum_{ij} \gamma_{ij}^\dagger \gamma_{ij} = \sigma^{-\frac{1}{2}}\tr_C \left(\rowE*{\sigma} \rowE{\sigma}\right) \sigma^{-\frac{1}{2}} = \sigma^{-\frac{1}{2}} \sigma \sigma^{-\frac{1}{2}} = I.
\end{equation}
Therefore, as long as $\sigma$ is full rank, the composition of the Petz recovery map and the encoding channel is CPTP.

\section{Near-Optimality of the Petz Map}\label{sec:near_optimality}
Consider the entanglement fidelity of a general recovery channel $\R$ for the encoding channel $\E$, evaluated at the input state $\tau=\sigma$. In graphical notation, it can be represented as
\begin{equation}\label{eq:ent_fidelity_diagram}
    \begin{aligned}
    \Fent_\sigma(\R \circ \E) = \begin{tikzpicture}[baseline={([yshift=-0.5ex]current bounding box.center)}]
        % Equal tensor-box heights keep corresponding wire endpoints aligned.
        \begin{scope}[every node/.append style={text height=2.2ex, text depth=0.8ex}]
        \pic[kraus/label = $\colE{I}$] (E1) at (0,0) {kraus};
        \pic[kraus/label = $\colE*{I}$] (E2) at (0,2) {krausd};
        \pic[kraus/label = $R^T$] (R1) at (2,0) {kraust};
        \pic[kraus/label = $R^*$] (R2) at (2,2) {krausstar};
        \end{scope}
        \draw[thick] (E1-A) -- (R1-D) node[midway, rectangle, draw=black, fill=white!5, very thick, minimum size=.2cm] {$\sigma$};
        \draw[thick] (E2-A) -- (R2-D) node[midway, rectangle, draw=black, fill=white!5, very thick, minimum size=.2cm] {$\sigma$};
        \draw[thick] (E1-D) -- (R1-A);
        \draw[thick] (E2-D) -- (R2-A);
        \draw[thick] (E1-C) -- (E2-C);
        \draw[thick] (R1-C) --++ (0.5, 0) |- (R2-C);    
    \end{tikzpicture}.
    \end{aligned}
\end{equation}
Using Eq.~\eqref{eq:Esigma_identity}, we can rewrite the entanglement fidelity as
\begin{equation}
    \begin{tikzpicture}[baseline={([yshift=-0.5ex]current bounding box.center)}]
        \pic[kraus/label = $\colE{I}$] (E1) at (0,0) {kraus};
        \pic[kraus/label = $\colE*{I}$] (E2) at (0,3) {krausd};
        \pic[kraus/label = $R^T$] (R1) at (5,0) {kraust};
        \pic[kraus/label = $R^*$] (R2) at (5,3) {krausstar};
        \draw[thick] (E1-A) --++(0, -0.25) -| (R1-D) node[pos=0.1, rectangle, draw=black, fill=white!5, very thick, minimum size=.2cm] {$\sigma^{3/4}$} node[pos=0.4, rectangle, draw=black, fill=white!5, very thick, minimum size=.2cm] {$\sigma^{1/4}$};
        \draw[thick] (E2-A) --++(0, 0.25) -| (R2-D) node[pos=0.1, rectangle, draw=black, fill=white!5, very thick, minimum size=.2cm] {$\sigma^{3/4}$} node[pos=0.4, rectangle, draw=black, fill=white!5, very thick, minimum size=.2cm] {$\sigma^{1/4}$};
        \draw[thick] (E1-D) --++(0, 0.25) -| (R1-A) node[pos=0.1, rectangle, draw=black, fill=white!5, very thick, minimum size=.2cm] {$\E(\sigma)^{-\frac{1}{4}}$} node[pos=0.4, rectangle, draw=black, fill=white!5, very thick, minimum size=.2cm] {$\E(\sigma)^{\frac{1}{4}}$};
        \draw[thick] (E2-D) --++(0, -0.25) -| (R2-A) node[pos=0.1, rectangle, draw=black, fill=white!5, very thick, minimum size=.2cm] {$\E(\sigma)^{-\frac{1}{4}}$} node[pos=0.4, rectangle, draw=black, fill=white!5, very thick, minimum size=.2cm] {$\E(\sigma)^{\frac{1}{4}}$};
        \draw[thick] (E1-C) -- (E2-C);
        \draw[thick] (R1-C) --++ (0, -0.25)--++ (0.5, 0) |- ([yshift=0.25cm]R2-C) -- (R2-C);
        \draw[dashed, draw=red] (2.5,-1.5) -- (2.5,4.5);    
    \end{tikzpicture}.
\end{equation}
Then we can split the diagram along the red dashed line using the Cauchy--Schwarz inequality:
\begin{equation}
    \Fent_\sigma (\R \circ \E)^2 \leq F_{L} F_{R},
\end{equation}
where $F_{L}$ is
\begin{equation}
    \begin{tikzpicture}
        \begin{scope}[every node/.append style={text height=2.2ex, text depth=0.8ex}]
        \pic[kraus/label = $\colE{I}$] (E1) at (0,0) {kraus};
        \pic[kraus/label = $\colE*{I}$] (E2) at (0,3) {krausd};
        \pic[kraus/label = ${\colE{I}}^*$] (R1) at (3,0) {krausstar};
        \pic[kraus/label = ${\colE{I}}^T$] (R2) at (3,3) {kraust};
        \end{scope}
        \draw[thick] (E1-A) --++(0, -0.25) -- ([yshift=-0.25cm]R1-A) node[pos=0.5, rectangle, draw=black, fill=white!5, very thick, minimum size=.2cm] {$\sigma^{3/2}$} -- (R1-A);
        \draw[thick] (E2-A) --++(0, 0.25) -- ([yshift=0.25cm]R2-A) node[pos=0.5, rectangle, draw=black, fill=white!5, very thick, minimum size=.2cm] {$\sigma^{3/2}$} -- (R2-A);
        \draw[thick] (E1-C) -- (E2-C);
        \draw[thick] (R1-C) -- (R2-C);
        \draw[thick] (E1-D) --++ (0, 0.25) -- ([yshift=0.25cm]R1-D)  node[pos=0.5, rectangle, draw=black, fill=white!5, very thick, minimum size=.2cm] {$\E(\sigma)^{-1/2}$} -- (R1-D);
        \draw[thick] (E2-D) --++ (0, -0.25) -- ([yshift=-0.25cm]R2-D)  node[pos=0.5, rectangle, draw=black, fill=white!5, very thick, minimum size=.2cm] {$\E(\sigma)^{-1/2}$} -- (R2-D);
    \end{tikzpicture}.
\end{equation}
This is precisely $\Fent_\sigma (\petzChannel{\sigma} \circ \E)$, the entanglement fidelity of the Petz recovery map. The right factor $F_{R}$ is
\begin{equation}
    \begin{tikzpicture}
        \pic[kraus/label = $R$] (E1) at (0,0) {kraus};
        \pic[kraus/label = $R^\dagger$] (E2) at (0,3) {krausd};
        \pic[kraus/label = $R^*$] (R1) at (3,0) {krausstar};
        \pic[kraus/label = $R^T$] (R2) at (3,3) {kraust};
        \draw[thick] (E1-A) --++(0, -0.25) -- ([yshift=-0.25cm]R1-A) node[pos=0.5, rectangle, draw=black, fill=white!5, very thick, minimum size=.2cm] {$\E(\sigma)^{1/2}$} -- (R1-A);
        \draw[thick] (E2-A) --++(0, 0.25) -- ([yshift=0.25cm]R2-A) node[pos=0.5, rectangle, draw=black, fill=white!5, very thick, minimum size=.2cm] {$\E(\sigma)^{1/2}$} -- (R2-A);
        \draw[thick] (E1-C) -- (E2-C);
        \draw[thick] (R1-C) -- (R2-C);
        \draw[thick] (E1-D) --++ (0, 0.25) -- ([yshift=0.25cm]R1-D)  node[pos=0.5, rectangle, draw=black, fill=white!5, very thick, minimum size=.2cm] {$\sigma^\frac{1}{2}$} -- (R1-D);
        \draw[thick] (E2-D) --++ (0, -0.25) -- ([yshift=-0.25cm]R2-D)  node[pos=0.5, rectangle, draw=black, fill=white!5, very thick, minimum size=.2cm] {$\sigma^\frac{1}{2}$} -- (R2-D);
    \end{tikzpicture}.
\end{equation}
The vectorizations of $\E(\sigma)^{1/2}$ and $\sigma^{1/2}$ are normalized pure states. The diagram gives the overlap of the latter with the output of the trace-preserving channel $\R\otimes\R^*$ acting on the former, where $\R^*$ has Kraus operators $R_i^*$. Hence $F_R\leq 1$. Combining everything leads to the near-optimality inequality for the Petz recovery map:
\begin{equation}
    \Fent_\sigma (\petzChannel{\sigma} \circ \E) \geq  \Fent_\sigma (\R \circ \E)^2.
\end{equation}
The recovery channel $\R$ on the right-hand side of the inequality is arbitrary and can be the optimal recovery channel that maximizes the entanglement fidelity.

\section{Bounding the Diamond Distance}
\label{sec:diamond_distance}
The diamond distance measures the worst-case distinguishability of two channels, allowing inputs entangled with an ancillary system.

% The diamond distance between two quantum channels quantifies the distinguishability between them, capturing the maximum difference in their action on any input state, including entangled states with an ancillary system.
% We now study the diamond distance between the Petz map and its approximation, aiming to understand how closely the approximate map can mimic the action of the exact Petz recovery map at a given number of iterations $k$ and angle sequence $\{\theta_i\}$. The unextended Petz map vanishes on states supported in the null space of $\E(\sigma)$, whereas a trace-preserving extension must specify an action there. For full-rank $\sigma$, comparing the composite encoding--recovery channels avoids this extension-dependent behavior. We therefore compare $\R^{\vec \theta} \circ \E$ and $\petzChannel{\sigma} \circ \E$, which map states from $\H_A$ to $\H_A$.
For full-rank $\sigma$, every encoded state is supported on $\operatorname{supp}\E(\sigma)$. We therefore compare the composite channels $\R^{\vec\theta}\circ\E$ and $\petzChannel{\sigma}\circ\E$, whose actions are independent of how Petz recovery is extended outside this support.

In the composite channel, we label the encoding environment by $C$ and the recovery environment by $C'$, a copy of the $C$ register used in the standalone recovery construction. Define the recovery isometries
\[
    T^{\vec\theta}=\recoveryW\ket{s},\qquad
    T^{\petz}_\sigma=\petzW{\sigma}\ket{s}.
\]
The corresponding composite encoding--recovery isometries are
\begin{equation}
    \begin{aligned}
    V^{\vec{\theta}} &= \left(I_C\otimes T^{\vec\theta}\right) \colE{I},\\
    V^{\petz}_\sigma &= \left(I_C\otimes T^{\petz}_\sigma\right) \colE{I},
    \end{aligned}
\end{equation}
each mapping $\H_A$ to $\H_C\otimes\H_T\otimes\H_{C'}\otimes\H_A$, with dimensions $d_C d_T d_{C'}d_A \times d_A$ and $d_{C'}=d_C$. The composite channels can then be implemented as
\begin{equation}
    \R^{\vec \theta} \circ \E(\rho)  = \tr_{C\otimes T\otimes C'} \left[ V^{\vec \theta} \rho V^{\vec \theta \dagger} \right],
\end{equation}
and similarly for the Petz recovery map.

Since the diamond distance decreases under common post-processing, we have
\begin{equation}
    \|\R^{\vec \theta} \circ \E - \petzChannel{\sigma} \circ \E\|_\diamond \leq \|\mathcal{V}^{\vec \theta} - \mathcal{V}^{\petz}_\sigma\|_\diamond,
\end{equation}
where $\mathcal{V}^{\vec \theta}(\rho) = V^{\vec \theta} \rho V^{\vec \theta \dagger}$ and $\mathcal{V}^{\petz}_\sigma(\rho) = V^{\petz}_\sigma \rho V^{\petz \dagger}_\sigma$. This lets us bound the diamond distance between the composite channels by the diamond distance between the corresponding isometry channels.
The diamond distance between two isometry channels is given by
\begin{equation}
    \frac{1}{2}\|\mathcal{V}^{\vec \theta} - \mathcal{V}^{\petz}_\sigma\|_\diamond =  \max _\rho \sqrt{1 - |\tr (\rho V^{\vec \theta \dagger} V^{\petz}_\sigma)|^2},
\end{equation}
where $\rho$ is a density matrix in $\H_A$. Let $f(\rho) = |\tr (\rho V^{\vec \theta \dagger} V^{\petz}_\sigma)|$. The optimal solution $\rho^*$ minimizes $f(\rho)$. We then establish the following two inequalities:
\begin{equation}
    1 + f(\rho^*) \leq 1 + f(\rho), \quad (1-f(\rho^*))\leq \|\rho^{-1}\|_{\infty} (1-f(\rho)),
\end{equation}
where $\rho$ is an arbitrary full-rank density matrix, and $\|\rho^{-1}\|_{\infty}$ is the inverse of the smallest eigenvalue of $\rho$. The first inequality follows from the optimality of $\rho^*$. For the second, set $\mu=\lambda_{\min}(\rho)$. Since $\rho\geq\mu I\geq\mu\rho^*$, we can write $\rho=\mu\rho^*+(1-\mu)\tau$ for a density matrix $\tau$ when $\mu<1$. The isometry condition gives $\|V^{\vec\theta\dagger}V^{\petz}_\sigma\|_\infty\leq1$, so the triangle inequality implies
\[
f(\rho)\leq\mu f(\rho^*)+(1-\mu)f(\tau)
\leq\mu f(\rho^*)+1-\mu.
\]
Rearranging proves the second inequality; the case $\mu=1$ is immediate. Multiplying the two inequalities leads to
\begin{equation}
    1-f(\rho^*)^2 \leq \|\rho^{-1}\|_{\infty}(1-f(\rho)^2).
\end{equation}
To evaluate $f(\sigma)$, we trace out the encoding environment:
\[
\begin{aligned}
\tr\!\left(\sigma V^{\vec\theta\dagger}V^{\petz}_\sigma\right)
&=\tr\!\left[\E(\sigma)(T^{\vec\theta})^\dagger T^{\petz}_\sigma\right]\\
&=\sum_i\lambda_i^2 p(x_i)^*.
\end{aligned}
\]
Here $\E(\sigma)=\sum_i\lambda_i^2\ket{E_i^{D}}\bra{E_i^{D}}$, and the error component in Eq.~\eqref{eq:W_qsvt} is orthogonal to the target auxiliary state $\ket{t}$. Taking the absolute value removes the complex conjugation and yields
\begin{equation}
    f(\sigma) = \left|\sum_i \lambda_i^2 p(x_i)\right|,
\end{equation}
where $\sigma$ is the full-rank reference state used in the Petz map, $\lambda_i$ are the singular values of $\rowE{\sigma}$, $x_i=\lambda_i/\alpha$ are the scaled singular values, and $p$ is an odd polynomial of degree at most $2k+1$ from the QSVT. Combining everything, we obtain a simple bound on the diamond distance in terms of the polynomial $p$ and the singular values $\lambda$:
\begin{equation}
    \Delta(\vec\theta) = \frac{1}{2}\|\R^{\vec \theta} \circ \E - \petzChannel{\sigma} \circ \E\|_\diamond \leq \sqrt{\|\sigma^{-1}\|_{\infty} \mathcal{I}^{\vec{\theta}}},
\end{equation}
where 
\begin{equation}
    \mathcal{I}^{\vec{\theta}} = 1 - \left|\sum_i \lambda_i^2 p(x_i)\right|^2.
\end{equation}
Clearly, the diamond distance between the composite channels is controlled by the polynomial $p$ and the singular value distribution of $\rowE{\sigma}$. If $p(x)$ is the sign function apart from a global phase, then the upper bound is zero.

\section{The Entanglement Fidelity of the Approximate Petz Channel}\label{sec:approximate_petz_fidelity}
Here we analyze the entanglement fidelity of the approximate Petz channel $\R^{\vec \theta} \circ \E$, with recovery isometry $T^{\vec\theta}=\recoveryW\ket{s}$ as defined in Appendix~\ref{sec:diamond_distance}. For a general logical reference state $\sigma$, the diagrammatic representation is
\begin{equation}
    \begin{aligned}
    \Fent_\sigma(\R^{\vec\theta} \circ \E) = \begin{tikzpicture}[baseline={([yshift=-0.5ex]current bounding box.center)}]
        % Equal tensor-box heights keep corresponding wire endpoints aligned.
        \begin{scope}[every node/.append style={text height=2.2ex, text depth=0.8ex}]
        \pic[kraus/label = $\colE{I}$] (E1) at (0,0) {kraus};
        \pic[kraus/label = $\colE*{I}$] (E2) at (0,2) {krausd};
        \pic[kraus/label = $(T^{\vec\theta})^T$] (R1) at (2,0) {kraust};
        \pic[kraus/label = $(T^{\vec\theta})^*$] (R2) at (2,2) {krausstar};
        \end{scope}
        \draw[thick] (E1-A) -- (R1-D) node[midway, rectangle, draw=black, fill=white!5, very thick, minimum size=.2cm] {$\sigma$};
        \draw[thick] (E2-A) -- (R2-D) node[midway, rectangle, draw=black, fill=white!5, very thick, minimum size=.2cm] {$\sigma$};
        \draw[thick] (E1-D) -- (R1-A);
        \draw[thick] (E2-D) -- (R2-A);
        \draw[thick] (E1-C) -- (E2-C);
        \draw[thick] (R1-C) --++ (0.5, 0) |- (R2-C);    
    \end{tikzpicture}.
    \end{aligned}
\end{equation}
Below we specialize to the maximally mixed logical input, for which the EPR contractions become normalized partial traces:
\begin{equation}
    \Fent(\R^{\vec\theta}\circ\E)
    =\left\|\frac{1}{d_A}\tr_A\!\left[T^{\vec\theta}\colE{I}\right]\right\|_2^2.
\end{equation}
Here $\tr_A$ contracts the input and recovered logical indices, leaving the environment indices open, and $\|X\|_2^2=\tr(X^\dagger X)$ is the squared Hilbert--Schmidt norm of the remaining tensor. Identities on untouched environments are implicit. The reference state $\sigma$ defining the recovery need not be maximally mixed.

Recall that
\begin{equation}
    \begin{aligned}
    T^{\vec\theta} = \recoveryW\ket{s} &= \ket{t} P \\ &+ (1-\ket{t}\bra{t}) W\ket{s} Q.
    \end{aligned}
\end{equation}
The two terms are orthogonal on $T$. Expanding the complementary projector therefore gives
\begin{equation}
    \begin{aligned}
    \Fent ={}&\frac{1}{d_A^2}\left\|\tr_A\!\left[P\colE{I}\right]\right\|_2^2\\
    &-\frac{1}{d_A^2}\left\|\tr_A\!\left[Q'\colE{I}\right]\right\|_2^2\\
    &+\frac{1}{d_A^2}\left\|\tr_A\!\left[W\ket{s}Q\colE{I}\right]\right\|_2^2,
    \end{aligned}
\end{equation}
where $Q' = \bra{t} W\ket{s} Q$.
The first two terms depend only on the block-encoded operator and the chosen polynomials, while the last term depends on the full isometry $W\ket{s}$. For block encodings that prepare $\ket{\sigma}_{AA'}$ and leave its output half $A$ untouched by $W$, including generic types I and II in Eqs.~\eqref{eq:block_encoding_form1} and \eqref{eq:block_encoding_form2} and reduced type (II$'$)-HP in Eq.~\eqref{eq:block_encoding_form2_reduced_hp}, tracing over the other outputs leaves $\sigma_A$ independently of the input. Consequently,
\begin{equation}
    \begin{aligned}
    \left\|\tr_A\!\left[W\ket{s}Q\colE{I}\right]\right\|_2^2
    &=\tr\!\left[Q\E(\sigma)Q^\dagger\right]\\
    &=\sum_i\lambda_i^2|q(x_i)|^2.
    \end{aligned}
\end{equation}
For these block encodings, the entanglement fidelity simplifies to
\begin{equation}
    \begin{aligned}
    \Fent ={}&\frac{1}{d_A^2}\left\|\tr_A\!\left[P\colE{I}\right]\right\|_2^2\\
    &-\frac{1}{d_A^2}\left\|\tr_A\!\left[Q'\colE{I}\right]\right\|_2^2
    +\frac{1}{d_A^2}\sum_i\lambda_i^2|q(x_i)|^2.
    \end{aligned}
\end{equation}
Type I$'$ in Eq.~\eqref{eq:block_encoding_form1_mixed} and its HP version in Eq.~\eqref{eq:block_encoding_form1_mixed_hp} do not satisfy the untouched-output condition; their fidelities are given by the general three-term expression above.

\section{Estimation of the Singular-Value Distribution}\label{sec:sv}

In this appendix, we estimate the distribution of the nonzero singular values $\lambda$ of $\rowE{\sigma}$ for both the Haar-random and HP encoding channels. An overline denotes the uniform average over the nonzero singular values for a fixed encoding unitary, while $\mathbb{E}_U$ denotes the Haar average over encoding unitaries. Since $\E(\sigma) = \rowE{\sigma} \rowE*{\sigma}$ is a density matrix, $\sum_i\lambda_i^2=1$. Thus, with $r=\operatorname{rank}\E(\sigma)$ and $x_i=\lambda_i/\alpha$,
\[
    \overline{\lambda^2}=\frac{1}{r},
    \qquad \overline{x^2}=\frac{1}{r\alpha^2}.
\]
We define $\Std(x^2)$ as the square root of the Haar-averaged spectral variance. For fixed $\alpha$ and a rank $r$ that is constant almost surely, the identity $\tr[\E(\sigma)^2]=\sum_i\lambda_i^4$ gives
\begin{equation}\label{eq:spectral_relative_width}
    \frac{\Std(x^2)}{\overline{x^2}}
    =\sqrt{r\,\mathbb{E}_U\!\left[\tr[\E(\sigma)^2]\right]-1}.
\end{equation}
The mean is therefore determined by the rank and block-encoding normalization, while the relative width depends only on the rank and average purity.

\subsection{Haar-Random Code}
For Haar-random encoding with full-rank $\sigma$, $r=\min(d_A d_C,d_D)$ almost surely. Choosing the smaller normalization factor of the two generic block encodings gives $\alpha^2=r$ and hence $\overline{x^2}=1/r^2$. We compute the average purity using Weingarten calculus~\cite{Collins2006Integration}, starting from its diagrammatic representation:
\begin{equation}
    \begin{tikzpicture}
        \pic (E1) at (0,0) {unitary};
        \draw ([xshift=-0.1cm]E1-B) --++(0.2,0);
        \pic (E2) at (0,1.5) {unitaryd};
        \draw ([xshift=-0.1cm]E2-B) --++(0.2,0);
        \pic (R1) at (2,0) {unitarystar};
        \draw ([xshift=-0.1cm]R1-B) --++(0.2,0);
        \pic (R2) at (2,1.5) {unitaryt};
        \draw ([xshift=-0.1cm]R2-B) --++(0.2,0);
        \draw[thick] (E1-A) --++(-0.5cm, 0) -- ([xshift=-0.5cm]E2-A) node[pos=0.5, rectangle, draw=black, fill=white!5, very thick, minimum size=.2cm] {$\sigma$} -- (E2-A);
        \draw[thick] (R1-A) --++(0.5cm,0) -- ([xshift=0.5cm]R2-A) node[pos=0.5, rectangle, draw=black, fill=white!5, very thick, minimum size=.2cm] {$\sigma$} -- (R2-A);
        \draw[thick] (E1-C) -- (E2-C);
        \draw[thick] (R1-C) -- (R2-C);
        \draw[thick] (E1-D) -- (R1-D);
        \draw[thick] (E2-D) -- (R2-D);
    \end{tikzpicture}.
\end{equation}
Weingarten calculus gives
\begin{equation}
\begin{aligned}
\mathbb{E}_U\!\left[\tr[\E(\sigma)^2]\right]
&=\frac{1}{d^2-1}\Bigl(d d_D \tr(\sigma^2)+ d d_C\\
&\qquad -d_D -d_C \tr(\sigma^2)\Bigr),
\end{aligned}
\end{equation}
where $d = d_Ad_B=d_C d_D$ is the dimension of the space on which $U$ acts. Substituting into Eq.~\eqref{eq:spectral_relative_width} and specializing to $\sigma=I_A/d_A$, we obtain
\begin{equation}
    \overline{x^2} = \frac{1}{(d_A d_C)^2}, \ \Std(x^2)=\overline{x^2}\frac{\sqrt{d d_A -1}}{d_D}\sqrt{\frac{d^2-d_D^2}{d^2-1}}
\end{equation}
for $d_A d_C \le d_D$, and
\begin{equation}
    \overline{x^2} = \frac{1}{d_D^2}, \ \Std(x^2)=\overline{x^2}\sqrt{\frac{d-d_A}{d_A}}\sqrt{\frac{d_D^2-1}{d^2-1}}
\end{equation}
for $d_A d_C > d_D$.
The spectral width vanishes at $d_D=1$ and $d_D=d$, and is comparable to the mean near $d_A d_C=d_D$.

\subsection{Hayden--Preskill Protocol}
For the HP encoding channel with full-rank $\sigma$, $r=\min(d_A d_C,d_Dd_B)$ almost surely, and the reduced block encoding has $\alpha^2=d_D/d_B$. Hence
\begin{equation}
    \overline{x^2} = \frac{1}{\min (d_A^2, d_D^2)},
\end{equation} 
which is $\mathcal{O}(1)$ if the number of logical qubits is $\mathcal{O}(1)$. For a single logical qubit and $n_D\geq 1$, $\overline{x^2} = 1/4$.

The diagram for the purity $\tr[\E(\sigma)^2]$ is similar to that for the Haar-random code:
\begin{equation}
    \begin{tikzpicture}
        \pic (E1) at (0,0) {unitary};
        \pic (E2) at (0,1.5) {unitaryd};
        \pic (R1) at (2,0) {unitarystar};
        \pic (R2) at (2,1.5) {unitaryt};
        \draw[thick] (E1-B) -- ++(0,-0.35cm) -| (R1-B);
        \node[rectangle, draw=black, fill=white!5, very thick, minimum size=.2cm] at ($ (E1-B)!0.5!(R1-B) + (0,-0.35cm) $) {$\frac{I_B}{d_B}$};
        \draw[thick] (E2-B) -- ++(0,0.35cm) -| (R2-B);
        \node[rectangle, draw=black, fill=white!5, very thick, minimum size=.2cm] at ($ (E2-B)!0.5!(R2-B) + (0,0.35cm) $) {$\frac{I_B}{d_B}$};
        \draw[thick] (E1-A) --++(-0.5cm, 0) -- ([xshift=-0.5cm]E2-A) node[pos=0.5, rectangle, draw=black, fill=white!5, very thick, minimum size=.2cm] {$\sigma$} -- (E2-A);
        \draw[thick] (R1-A) --++(0.5cm,0) -- ([xshift=0.5cm]R2-A) node[pos=0.5, rectangle, draw=black, fill=white!5, very thick, minimum size=.2cm] {$\sigma$} -- (R2-A);
        \draw[thick] (E1-C) -- (E2-C);
        \draw[thick] (R1-C) -- (R2-C);
        \draw[thick] (E1-D) -- (R1-D);
        \draw[thick] (E2-D) -- (R2-D);
    \end{tikzpicture}.
\end{equation}
Weingarten calculus gives
\begin{equation}
\begin{aligned}
\mathbb{E}_U\!\left[\tr[\E(\sigma)^2]\right]
&=\frac{1}{d^2-1}\Bigl(d d_D \tr(\sigma^2)+ \frac{d d_C}{d_B}\\
&\qquad -\frac{d_D}{d_B} -d_C \tr(\sigma^2)\Bigr).
\end{aligned}
\end{equation}
Substituting into Eq.~\eqref{eq:spectral_relative_width} and specializing to $\sigma=I_A/d_A$, we obtain the mean and standard deviation of $x_i^2$:
\begin{equation}
\overline{x^2} = \frac{1}{d_A^2}, \quad \Std(x^2) = \frac{1}{d_A^2}\frac{\sqrt{d_A^2-1}}{d_D}\sqrt\frac{d^2-d_D^2}{d^2-1}
\end{equation}
for $d_A \leq d_D$, and
\begin{equation}
    \overline{x^2} = \frac{1}{d_D^2}, \quad \Std(x^2) = \frac{1}{d_D^2}\frac{\sqrt{d_D^2-1}}{d_A}\sqrt\frac{d^2-d_A^2}{d^2-1}
\end{equation}
for $d_A > d_D$.
Therefore, the relative spectral width $\Std(x^2)/\overline{x^2}$ is largest when $d_D=d_A$.

\end{document}